\documentclass[longauth]{aa}  

\usepackage{graphicx}
\usepackage{txfonts,textcomp}
\usepackage[colorlinks=true, allcolors=blue]{hyperref}
\usepackage{multirow}
\usepackage{placeins}

\usepackage{orcidlink}

\newcommand{\blz}{{S4~0954+65}}

\newcommand{\eqb}{\begin{eqnarray}}
\newcommand{\eqe}{\end{eqnarray}}

\begin{document}

\title{IXPE observation of the low-synchrotron peaked blazar S4~0954+65 during an optical-X-ray flare}

\subtitle{}

\author{
Pouya M. Kouch \inst{\ref{UTU},\ref{FINCA}} \thanks{\href{mailto:pouya.kouch@utu.fi}{pouya.kouch@utu.fi}} \orcid{0000-0002-9328-2750}, 
Ioannis Liodakis \inst{\ref{AstroCrete},\ref{NASA_Alabama}} \orcid{0000-0001-9200-4006}, 
Francesco Fenu \inst{\ref{Rome_ASI}}, 
Haocheng Zhang \inst{\ref{UniMaryland2},\ref{NASA_Goddard}} \orcid{0000-0001-9826-1759}, 
Stella Boula \inst{\ref{INAF_Merate}} \orcid{0000-0001-7905-6928}, 
Riccardo Middei \inst{\ref{SSDC_Rome},\ref{INAF_Rome_Obs}} \orcid{0000-0001-9815-9092}, 
Laura Di Gesu \inst{\ref{Rome_ASI}} \orcid{0000-0002-5614-5028}, 
Georgios F. Paraschos \inst{\ref{MPI_Bonn}} \orcid{0000-0001-6757-3098}, 
Iv\'{a}n Agudo \inst{\ref{InstAstro_Granada}} \orcid{0000-0002-3777-6182}, 
Svetlana G. Jorstad \inst{\ref{BostonUni},\ref{StPetersburg}} \orcid{0000-0001-6158-1708}, 
Elina Lindfors \inst{\ref{UTU}} \orcid{0000-0002-9155-6199}, 
Alan P. Marscher \inst{\ref{BostonUni}} \orcid{0000-0001-7396-3332}, 
Henric Krawczynski \inst{\ref{StLouis}} \orcid{0000-0002-1084-6507}, 
Michela Negro \inst{\ref{LSU}} \orcid{0000-0002-6548-5622}, 
Kun Hu \inst{\ref{StLouis}}, 
Dawoon E. Kim \inst{\ref{INAF_Rome},\ref{UniRome1},\ref{UniRome2}} \orcid{0000-0001-5717-3736}, 
Elisabetta Cavazzuti \inst{\ref{Rome_ASI}} \orcid{0000-0001-7150-9638}, 
Manel Errando \inst{\ref{StLouis}} \orcid{0000-0002-1853-863X}, 
Dmitry Blinov \inst{\ref{AstroCrete},\ref{UniHeraklion}}, 
Anastasia Gourni \inst{\ref{UniHeraklion}}, 
Sebastian Kiehlmann \inst{\ref{AstroCrete},\ref{UniHeraklion}} \orcid{0000-0001-6314-9177}, 
Angelos Kourtidis \inst{\ref{UniHeraklion_math}}, 
Nikos Mandarakas \inst{\ref{AstroCrete},\ref{UniHeraklion}}, 
Nikolaos Triantafyllou \inst{\ref{UniHeraklion}}, 
Anna Vervelaki \inst{\ref{UniHeraklion}}, 
George A. Borman \inst{\ref{AstroCrimea}}, 
Evgenia N. Kopatskaya \inst{\ref{StPetersburg}}, 
Elena G. Larionova \inst{\ref{StPetersburg}}, 
Daria A. Morozova \inst{\ref{StPetersburg}}, 
Sergey S. Savchenko \inst{\ref{StPetersburg},\ref{StPetersburg_Pulkovo}}, 
Andrey A. Vasilyev \inst{\ref{StPetersburg}}, 
Ivan S. Troitskiy \inst{\ref{StPetersburg}} \orcid{0000-0002-4218-0148}, 
Tatiana S. Grishina \inst{\ref{StPetersburg}} \orcid{0000-0002-3953-6676}, 
Ekaterina V. Shishkina \inst{\ref{StPetersburg}} \orcid{0009-0002-2440-2947}, 
Alexey V. Zhovtan \inst{\ref{AstroCrimea}}, 
Francisco Jos\'e Aceituno \inst{\ref{InstAstro_Granada}} 
Giacomo Bonnoli \inst{\ref{INAF_Merate},\ref{InstAstro_Granada}} \orcid{0000-0003-2464-9077}, 
V\'{i}ctor Casanova \inst{\ref{InstAstro_Granada}}, 
Juan Escudero \inst{\ref{InstAstro_Granada}} \orcid{0000-0002-4131-655X}, 
Beatriz Ag\'{i}s-Gonz\'{a}lez \inst{\ref{AstroCrete}},  
C\'{e}sar Husillos \inst{\ref{geological_madrid},\ref{InstAstro_Granada}} \orcid{0000-0001-8286-5443}, 
Jorge Otero-Santos \inst{\ref{InstAstro_Granada}}, 
Vilppu Piirola \inst{\ref{UTU}} \orcid{0000-0003-0186-206X}, 
Alfredo Sota \inst{\ref{InstAstro_Granada}} \orcid{0000-0002-9404-6952}, 
Ioannis Myserlis \inst{\ref{IRAM},\ref{MPI_Bonn}} \orcid{0000-0003-3025-9497}, 
Mark Gurwell \inst{\ref{Harvard_Smithsonian}} \orcid{0000-0003-0685-3621}, 
Garrett K. Keating \inst{\ref{Harvard_Smithsonian}} \orcid{0000-0002-3490-146X}, 
Ramprasad Rao \inst{\ref{Harvard_Smithsonian}} \orcid{0000-0002-1407-7944}, 
Emmanouil Angelakis \inst{\ref{Angelakis}} \orcid{0000-0001-7327-5441}, 
Alexander Kraus \inst{\ref{MPI_Bonn}} \orcid{0000-0002-4184-9372}, 
%
Lucio Angelo Antonelli \inst{\ref{INAF_Rome_Obs},\ref{SSDC_Rome}} \orcid{0000-0002-5037-9034}, 
Matteo Bachetti \inst{\ref{INAF_Selargius}} \orcid{0000-0002-4576-9337}, 
Luca Baldini \inst{\ref{INFN_Pisa},\ref{UniPisa}} \orcid{0000-0002-9785-7726}, 
Wayne H. Baumgartner \inst{\ref{Washington_NavalResearch}} \orcid{0000-0002-5106-0463}, 
Ronaldo Bellazzini \inst{\ref{INFN_Pisa}} \orcid{0000-0002-2469-7063}, 
Stefano Bianchi \inst{\ref{UniRome3}} \orcid{0000-0002-4622-4240}, 
Stephen D. Bongiorno \inst{\ref{NASA_Alabama}} \orcid{0000-0002-0901-2097}, 
Raffaella Bonino \inst{\ref{INFN_Turin},\ref{UniTurin}} \orcid{0000-0002-4264-1215}, 
Alessandro Brez \inst{\ref{INFN_Pisa}} \orcid{0000-0002-9460-1821}, 
Niccol\`{o} Bucciantini \inst{\ref{INAF_Florence_Obs},\ref{UniFlorence},\ref{INAF_Florence}} \orcid{0000-0002-8848-1392}, 
Fiamma Capitanio \inst{\ref{INAF_Rome}} \orcid{0000-0002-6384-3027}, 
Simone Castellano \inst{\ref{INFN_Pisa}} \orcid{0000-0003-1111-4292}, 
Chien-Ting Chen \inst{\ref{USRA_Alabama}} \orcid{0000-0002-4945-5079}, 
Stefano Ciprini \inst{\ref{INFN_Rome},\ref{SSDC_Rome}} \orcid{0000-0002-0712-2479}, 
Enrico Costa \inst{\ref{INAF_Rome}} \orcid{0000-0003-4925-8523}, 
Alessandra De Rosa \inst{\ref{INAF_Rome}} \orcid{0000-0001-5668-6863}, 
Ettore Del Monte \inst{\ref{INAF_Rome}} \orcid{0000-0002-3013-6334}, 
Niccol\`{o} Di Lalla \inst{\ref{Stanford}} \orcid{0000-0002-7574-1298}, 
Alessandro Di Marco \inst{\ref{INAF_Rome}} \orcid{0000-0003-0331-3259}, 
Immacolata Donnarumma \inst{\ref{Rome_ASI}} \orcid{0000-0002-4700-4549}, 
Victor Doroshenko \inst{\ref{Tubingen}} \orcid{0000-0001-8162-1105}, 
Michal Dov\v{c}iak \inst{\ref{Astro_Prague}} \orcid{0000-0003-0079-1239}, 
Steven R. Ehlert \inst{\ref{NASA_Alabama}} \orcid{0000-0003-4420-2838}, 
Teruaki Enoto \inst{\ref{RIKEN}} \orcid{0000-0003-1244-3100}, 
Yuri Evangelista \inst{\ref{INAF_Rome}} \orcid{0000-0001-6096-6710}, 
Sergio Fabiani \inst{\ref{INAF_Rome}} \orcid{0000-0003-1533-0283}, 
Riccardo Ferrazzoli \inst{\ref{INAF_Rome}} \orcid{0000-0003-1074-8605}, 
Javier A. Garcia \inst{\ref{NASA_Goddard}} \orcid{0000-0003-3828-2448}, 
Shuichi Gunji \inst{\ref{Jap_Yamagata}} \orcid{0000-0002-5881-2445}, 
Kiyoshi Hayashida \inst{\ref{UniOsaka}}, 
Jeremy Heyl \inst{\ref{UBC}} \orcid{0000-0001-9739-367X}, 
Wataru Iwakiri \inst{\ref{Jap_Chiba}} \orcid{0000-0002-0207-9010}, 
Philip Kaaret \inst{\ref{NASA_Alabama}} \orcid{0000-0002-3638-0637}, 
Vladimir Karas \inst{\ref{Astro_Prague}} \orcid{0000-0002-5760-0459}, 
Fabian Kislat \inst{\ref{UNH}} \orcid{0000-0001-7477-0380}, 
Takao Kitaguchi \inst{\ref{RIKEN}}, 
Jeffery J. Kolodziejczak \inst{\ref{NASA_Alabama}} \orcid{0000-0002-0110-6136}, 
Fabio La Monaca \inst{\ref{INAF_Rome},\ref{UniRome2},\ref{UniRome1}} \orcid{0000-0001-8916-4156}, 
Luca Latronico \inst{\ref{INFN_Turin}} \orcid{0000-0002-0984-1856}, 
Simone Maldera \inst{\ref{INFN_Turin}} \orcid{0000-0002-0698-4421}, 
Alberto Manfreda \inst{\ref{INFN_Naples}} \orcid{0000-0002-0998-4953}, 
Fr\'{e}d\'{e}ric Marin \inst{\ref{Strasbourg}} \orcid{0000-0003-4952-0835}, 
Andrea Marinucci \inst{\ref{Rome_ASI}} \orcid{0000-0002-2055-4946}, 
Herman L. Marshall \inst{\ref{MIT}} \orcid{0000-0002-6492-1293}, 
Francesco Massaro \inst{\ref{INFN_Turin},\ref{UniTurin}} \orcid{0000-0002-1704-9850}, 
Giorgio Matt \inst{\ref{UniRome3}} \orcid{0000-0002-2152-0916}, 
Ikuyuki Mitsuishi \inst{\ref{Nagoya}}, 
Tsunefumi Mizuno \inst{\ref{Jap_UniHiroshima}} \orcid{0000-0001-7263-0296}, 
Fabio Muleri \inst{\ref{INAF_Rome}} \orcid{0000-0003-3331-3794}, 
Chi-Yung Ng \inst{\ref{HongKong}} \orcid{0000-0002-5847-2612}, 
Stephen L. O'Dell \inst{\ref{NASA_Alabama}} \orcid{0000-0002-1868-8056}, 
Nicola Omodei \inst{\ref{Stanford}} \orcid{0000-0002-5448-7577}, 
Chiara Oppedisano \inst{\ref{INFN_Turin}} \orcid{0000-0001-6194-4601}, 
Alessandro Papitto \inst{\ref{INAF_Rome_Obs}} \orcid{0000-0001-6289-7413}, 
George G. Pavlov \inst{\ref{PennState}} \orcid{0000-0002-7481-5259}, 
Abel Lawrence Peirson \inst{\ref{Stanford}} \orcid{0000-0001-6292-1911}, 
Matteo Perri \inst{\ref{SSDC_Rome},\ref{INAF_Rome_Obs}} \orcid{0000-0003-3613-4409}, 
Melissa Pesce-Rollins \inst{\ref{INFN_Pisa}} \orcid{0000-0003-1790-8018}, 
Pierre-Olivier Petrucci \inst{\ref{Grenoble}} \orcid{0000-0001-6061-3480}, 
Maura Pilia \inst{\ref{INAF_Selargius}} \orcid{0000-0001-7397-8091}, 
Andrea Possenti \inst{\ref{INAF_Selargius}} \orcid{0000-0001-5902-3731}, 
Juri Poutanen \inst{\ref{UTU}} \orcid{0000-0002-0983-0049}, 
Simonetta Puccetti \inst{\ref{SSDC_Rome}} \orcid{0000-0002-2734-7835}, 
Brian D. Ramsey \inst{\ref{NASA_Alabama}} \orcid{0000-0003-1548-1524}, 
John Rankin \inst{\ref{INAF_Merate}} \orcid{0000-0002-9774-0560}, 
Ajay Ratheesh \inst{\ref{INAF_Rome}} \orcid{0000-0003-0411-4243}, 
Oliver J. Roberts \inst{\ref{USRA_Alabama}} \orcid{0000-0002-7150-9061}, 
Carmelo Sgr\`{o} \inst{\ref{INFN_Pisa}} \orcid{0000-0001-5676-6214}, 
Patrick Slane \inst{\ref{Harvard_Smithsonian}} \orcid{0000-0002-6986-6756}, 
Paolo Soffitta \inst{\ref{INAF_Rome}} \orcid{0000-0002-7781-4104}, 
Gloria Spandre \inst{\ref{INFN_Pisa}} \orcid{0000-0003-0802-3453}, 
Douglas A. Swartz \inst{\ref{USRA_Alabama}} \orcid{0000-0002-2954-4461}, 
Toru Tamagawa \inst{\ref{RIKEN}} \orcid{0000-0002-8801-6263}, 
Fabrizio Tavecchio \inst{\ref{INAF_Merate}} \orcid{0000-0003-0256-0995}, 
Roberto Taverna \inst{\ref{UniPadova}} \orcid{0000-0002-1768-618X}, 
Yuzuru Tawara \inst{\ref{Nagoya}}, 
Allyn F. Tennant \inst{\ref{NASA_Alabama}} \orcid{0000-0002-9443-6774}, 
Nicholas E. Thomas \inst{\ref{NASA_Alabama}} \orcid{0000-0003-0411-4606}, 
Francesco Tombesi \inst{\ref{UniRome2},\ref{INFN_Rome},\ref{UniMaryland}} \orcid{0000-0002-6562-8654}, 
Alessio Trois \inst{\ref{INAF_Selargius}} \orcid{0000-0002-3180-6002}, 
Sergey S. Tsygankov \inst{\ref{UTU}} \orcid{0000-0002-9679-0793}, 
Roberto Turolla \inst{\ref{UniPadova},\ref{MSSL}} \orcid{0000-0003-3977-8760}, 
Roger W. Romani \inst{\ref{Stanford}} \orcid{0000-0001-6711-3286}, 
Jacco Vink \inst{\ref{Amsterdam}} \orcid{0000-0002-4708-4219}, 
Martin C. Weisskopf \inst{\ref{NASA_Alabama}} \orcid{0000-0002-5270-4240}, 
Kinwah Wu \inst{\ref{MSSL}} \orcid{0000-0002-7568-8765}, 
Fei Xie \inst{\ref{China_Guangxi},\ref{INAF_Rome}} \orcid{0000-0002-0105-5826}, 
Silvia Zane \inst{\ref{MSSL}} \orcid{0000-0001-5326-880X} 
}

\institute{
Department of Physics and Astronomy, University of Turku, FI-20014 Turku, Finland \label{UTU}
\and
Finnish Centre for Astronomy with ESO (FINCA), Quantum, Vesilinnantie 5, University of Turku, FI-20014 Turku, Finland \label{FINCA}
\and
Institute of Astrophysics, FORTH, N. Plastira 100, GR-70013 Heraklion, Greece \label{AstroCrete}
\and
NASA Marshall Space Flight Center, Huntsville, AL 35812, USA \label{NASA_Alabama}
\and
ASI - Agenzia Spaziale Italiana, Via del Politecnico snc, 00133 Roma, Italy \label{Rome_ASI}
\and
University of Maryland, Baltimore County, Baltimore, MD 21250, USA \label{UniMaryland2}
\and
NASA Goddard Space Flight Center, Greenbelt, MD 20771, USA \label{NASA_Goddard}
\and
INAF Osservatorio Astronomico di Brera, Via E. Bianchi 46, 23807 Merate (LC), Italy \label{INAF_Merate}
\and
Space Science Data Center, Agenzia Spaziale Italiana, Via del Politecnico snc, 00133 Roma, Italy \label{SSDC_Rome}
\and
INAF Osservatorio Astronomico di Roma, Via Frascati 33, 00078 Monte Porzio Catone (RM), Italy \label{INAF_Rome_Obs}
\and
Max-Planck-Institut f\"{u}r Radioastronomie, Auf dem H\"{u}gel 69, D-53121 Bonn, Germany \label{MPI_Bonn}
\and
Instituto de Astrof\'{i}sica de Andaluc\'{i}a, IAA-CSIC, Glorieta de la Astronom\'{i}a s/n, 18008 Granada, Spain \label{InstAstro_Granada}
\and
Institute for Astrophysical Research, Boston University, 725 Commonwealth Avenue, Boston, MA 02215, USA \label{BostonUni}
\and
Saint Petersburg State University, 7/9 Universitetskaya nab., St. Petersburg 199034, Russia \label{StPetersburg}
\and
Physics Department and McDonnell Center for the Space Sciences, Washington University in St. Louis, St. Louis, MO 63130, USA \label{StLouis}
\and
Department of Physics and Astronomy, Louisiana State University, Baton Rouge, LA 70803, USA \label{LSU}
\and
INAF Istituto di Astrofisica e Planetologia Spaziali, Via del Fosso del Cavaliere 100, 00133 Roma, Italy \label{INAF_Rome}
\and
Dipartimento di Fisica, Universit\`{a} degli Studi di Roma "La Sapienza", Piazzale Aldo Moro 5, 00185 Roma, Italy \label{UniRome1}
\and
Dipartimento di Fisica, Universit\`{a} degli Studi di Roma "Tor Vergata", Via della Ricerca Scientifica 1, 00133 Roma, Italy \label{UniRome2}
\and
Department of Physics, University of Crete, 70013, Heraklion, Greece \label{UniHeraklion}
\and
Department of Mathematics and Applied Mathematics - Applied Mathematics, University of Crete, 70013, Heraklion, Greece \label{UniHeraklion_math}
\and
Crimean Astrophysical Observatory RAS, P/O Nauchny, 298409, Crimea \label{AstroCrimea}
\and
Pulkovo Observatory, St. Petersburg, 196140, Russia \label{StPetersburg_Pulkovo}
\and
Geological and Mining Institute of Spain (IGME-CSIC), Calle R\'{i}os Rosas 23, E-28003, Madrid, Spain \label{geological_madrid}
\and
Institut de Radioastronomie Millim\'{e}trique, Avenida Divina Pastora, 7, Local 20, E–18012 Granada, Spain \label{IRAM}
\and
Center for Astrophysics | Harvard \& Smithsonian, 60 Garden St, Cambridge, MA 02138, USA \label{Harvard_Smithsonian}
\and
Orchideenweg 8, D-53123 Bonn, Germany \label{Angelakis}
\and
INAF Osservatorio Astronomico di Cagliari, Via della Scienza 5, 09047 Selargius (CA), Italy \label{INAF_Selargius}
\and
Istituto Nazionale di Fisica Nucleare, Sezione di Pisa, Largo B. Pontecorvo 3, 56127 Pisa, Italy \label{INFN_Pisa}
\and
Dipartimento di Fisica, Universit\`{a} di Pisa, Largo B. Pontecorvo 3, 56127 Pisa, Italy \label{UniPisa}
\and
Naval Research Laboratory, 4555 Overlook Ave. SW, Washington, DC 20375, USA \label{Washington_NavalResearch}
\and
Dipartimento di Matematica e Fisica, Universit\`{a} degli Studi Roma Tre, Via della Vasca Navale 84, 00146 Roma, Italy \label{UniRome3}
\and
Istituto Nazionale di Fisica Nucleare, Sezione di Torino, Via Pietro Giuria 1, 10125 Torino, Italy \label{INFN_Turin}
\and
Dipartimento di Fisica, Universit\`{a} degli Studi di Torino, Via Pietro Giuria 1, 10125 Torino, Italy \label{UniTurin}
\and
INAF Osservatorio Astrofisico di Arcetri, Largo Enrico Fermi 5, 50125 Firenze, Italy \label{INAF_Florence_Obs}
\and
Dipartimento di Fisica e Astronomia, Universit\`{a} degli Studi di Firenze, Via Sansone 1, 50019 Sesto Fiorentino (FI), Italy \label{UniFlorence}
\and
Istituto Nazionale di Fisica Nucleare, Sezione di Firenze, Via Sansone 1, 50019 Sesto Fiorentino (FI), Italy \label{INAF_Florence}
\and
Science and Technology Institute, Universities Space Research Association, Huntsville, AL 35805, USA \label{USRA_Alabama}
\and
Istituto Nazionale di Fisica Nucleare, Sezione di Roma "Tor Vergata", Via della Ricerca Scientifica 1, 00133 Roma, Italy \label{INFN_Rome}
\and
Department of Physics and Kavli Institute for Particle Astrophysics and Cosmology, Stanford University, Stanford, California 94305, USA \label{Stanford}
\and
Institut f\"{u}r Astronomie und Astrophysik, Universit\"{a}t T\"{u}bingen, Sand 1, 72076 T\"{u}bingen, Germany \label{Tubingen}
\and
Astronomical Institute of the Czech Academy of Sciences, Bo\v{c}n\'{i} II 1401/1, 14100 Praha 4, Czech Republic \label{Astro_Prague}
\and
RIKEN Cluster for Pioneering Research, 2-1 Hirosawa, Wako, Saitama 351-0198, Japan \label{RIKEN}
\and
Yamagata University, 1-4-12 Kojirakawa-machi, Yamagata-shi 990-8560, Japan \label{Jap_Yamagata}
\and
Osaka University, 1-1 Yamadaoka, Suita, Osaka 565-0871, Japan \label{UniOsaka}
\and
University of British Columbia, Vancouver, BC V6T 1Z4, Canada \label{UBC}
\and
International Center for Hadron Astrophysics, Chiba University, Chiba 263-8522, Japan \label{Jap_Chiba}
\and
Department of Physics and Astronomy and Space Science Center, University of New Hampshire, Durham, NH 03824, USA \label{UNH}
\and
Istituto Nazionale di Fisica Nucleare, Sezione di Napoli, Strada Comunale Cinthia, 80126 Napoli, Italy \label{INFN_Naples}
\and
Universit\'{e} de Strasbourg, CNRS, Observatoire Astronomique de Strasbourg, UMR 7550, 67000 Strasbourg, France \label{Strasbourg}
\and
MIT Kavli Institute for Astrophysics and Space Research, Massachusetts Institute of Technology, 77 Massachusetts Avenue, Cambridge, MA 02139, USA \label{MIT}
\and
Graduate School of Science, Division of Particle and Astrophysical Science, Nagoya University, Furo-cho, Chikusa-ku, Nagoya, Aichi 464-8602, Japan \label{Nagoya}
\and
Hiroshima Astrophysical Science Center, Hiroshima University, 1-3-1 Kagamiyama, Higashi-Hiroshima, Hiroshima 739-8526, Japan \label{Jap_UniHiroshima}
\and
Department of Physics, The University of Hong Kong, Pokfulam, Hong Kong \label{HongKong}
\and
Department of Astronomy and Astrophysics, Pennsylvania State University, University Park, PA 16802, USA \label{PennState}
\and
Universit\'{e} Grenoble Alpes, CNRS, IPAG, 38000 Grenoble, France \label{Grenoble}
\and
Dipartimento di Fisica e Astronomia, Universit\`{a} degli Studi di Padova, Via Marzolo 8, 35131 Padova, Italy \label{UniPadova}
\and
Department of Astronomy, University of Maryland, College Park, Maryland 20742, USA \label{UniMaryland}
\and
Mullard Space Science Laboratory, University College London, Holmbury St Mary, Dorking, Surrey RH5 6NT, UK \label{MSSL}
\and
Anton Pannekoek Institute for Astronomy \& GRAPPA, University of Amsterdam, Science Park 904, 1098 XH Amsterdam, The Netherlands \label{Amsterdam}
\and
Guangxi Key Laboratory for Relativistic Astrophysics, School of Physical Science and Technology, Guangxi University, Nanning 530004, China \label{China_Guangxi}
}


\date{Received November 22, 2024; accepted January 31, 2025}

 
\abstract{The X-ray polarization observations, made possible with the Imaging X-ray Polarimetry Explorer (IXPE), offer new ways of probing high-energy emission processes in astrophysical jets from blazars. Here, we report the first X-ray polarization observation of the blazar S4 0954+65 in a high optical and X-ray state. During our multi-wavelength (MWL) campaign of the source, we detected an optical flare whose peak coincided with the peak of an X-ray flare. This optical-X-ray flare most likely took place in a feature moving along the parsec-scale jet, imaged at 43~GHz by the Very Long Baseline Array (VLBA). The 43~GHz polarization angle of the moving component underwent a rotation near the time of the flare. In the optical band, prior to the IXPE observation, we measured the polarization angle to be aligned with the jet axis. In contrast, during the optical flare, the optical polarization angle was perpendicular to the jet axis; after the flare, it reverted to being parallel to the jet axis. Due to the smooth behavior of the optical polarization angle during the flare, we favor shocks as the main acceleration mechanism. We also infer that the ambient magnetic field lines in the jet were parallel to the jet position angle. The average degree of optical polarization during the IXPE observation was (14.3$\pm$4.1)\%. Despite the flare, we only detected an upper limit of 14\% (at 3$\sigma$ level) on the X-ray polarization degree; however, a reasonable assumption on the X-ray polarization angle results in an upper limit of 8.8\% ($3\sigma$). We modeled the spectral energy distribution (SED) and spectral polarization distribution (SPD) of S4 0954+65 with leptonic (synchrotron self-Compton) and hadronic (proton and pair synchrotron) models. Our combined MWL polarization observations and SED modeling tentatively disfavor the use of hadronic models for the X-ray emission in S4 0954+65.}


\titlerunning{IXPE observation of the LSP blazar S4 0954+65}
\authorrunning{Kouch et al.}

\maketitle
%
\section{Introduction} \label{sec_intro}


Blazars are active galactic nuclei (AGNs) with relativistic jets aligned within a small viewing angle toward an observer. These powerful objects are observed through the entire electromagnetic spectrum from radio to very-high-energy (VHE; 100~GeV to 100~TeV) $\gamma$-rays. Their energy output at almost all wavelengths exhibits rapid, seemingly stochastic variations (e.g., \citealt{Ulrich1997_blz_var}). Conventionally, blazars are classified into two subgroups: BL Lac objects (BLLs), with largely featureless optical spectra, and flat spectrum radio quasars (FSRQs), which exhibit strong emission lines. Spectral energy distributions (SEDs) of blazars typically exhibit a double-humped structure. It is generally accepted that the low-energy component of a blazar SED arises due to the synchrotron emission of electrons and positrons radiating in the magnetized jet. Based on the peak frequency of this synchrotron component ($\nu_{\mathrm{Sy}}$), blazars have been further subdivided into low-, intermediate-, and high-synchrotron peaked sources (e.g., \citealt{abdo2010}): LSP ($\nu_{\mathrm{Sy}}$<$10^{14}$~Hz), ISP ($10^{14}$<$\nu_{\mathrm{Sy}}$<$10^{15}$~Hz), and HSP ($\nu_{\mathrm{Sy}}$>$10^{15}$~Hz). For recent reviews of blazar studies, we refer to  \cite{Bottcher19}, \cite{blandford2019}, and \cite{hovatta_lindfors2019}.

The origin of the high-energy component of blazar SEDs is still not fully understood. Possibilities include the commonly adopted leptonic models, which explain the high-energy emission as Compton scattering. In these models, the same electrons whose synchrotron emission produces the low-energy SED component upscatter their own synchrotron photons or photons from external sources to higher energies. The former models are referred to as synchrotron self-Compton (SSC; e.g., \citealt{jones1974,maraschi1992}), while the latter are referred to as external Compton (EC; e.g., \citealt{dermer1992,sikora1994}). On the other hand, hadronic emission models invoke synchrotron emission by protons or secondary particles produced in photo-pion cascades, to explain the origin of the high-energy component of blazar SEDs (e.g., \citealt{mannheim1989}). Hadronic emission models are typically less favored than leptonic ones (e.g., \citealt{Bottcher2013}), since, for them to be efficient, the total energy requirement in protons is often unsustainably large. It can sometimes be even higher than the Eddington luminosity, since a large fraction of protons must reach energies $\gtrsim \mathrm{PeV}$ to exceed the photo-pion energy threshold (e.g., \citealt{Boettcher2022}). Additionally, hadronic models require rather high magnetic field strengths ($\sim$10--300~G, e.g., \citealt{Liodakis2020}). Nevertheless, the recent $\sim$3$\sigma$ spatio-temporal association of the blazar TXS~0506+056 with high-energy neutrinos detected by the IceCube Neutrino Observatory (\citealt{ic2018}), coupled with the possibility that there may be a population-based correlation between high-energy neutrinos and blazars (or at least a subpopulation of blazars; e.g., \citealt{padovani2016_HSPs_as_neut_emitters,Kouch2024_CGRaBS_update}), suggests that hadronic processes may be important in blazar jets (e.g., \citealt{Petropoulou2020}).


Given the inherent differences in the polarization properties of electron synchrotron, proton synchrotron, and Compton radiation, multi-wavelength (MWL) polarization measurements are crucial for distinguishing between different particle acceleration and emission models for blazar jets (e.g., \citealt{Marscher2008_nature_MW_EVPA_rot}). Linear polarization is typically represented via two quantities derived from the Stokes parameters (Q, U, and I)\footnote{Another Stokes parameter V measures circular polarization, which is intrinsically small in blazars ($\lesssim1\%$; \citealt{Liodakis2022circular,Liodakis2023_circ_PD}). It is usually ignored and not measured by IXPE at all. When including V, the polarization degree is $\Pi=\sqrt{\mathrm{Q}^2+\mathrm{U}^2+\mathrm{V}^2}/\mathrm{I}$).}: the polarization degree ($\Pi=\sqrt{\mathrm{Q}^2+\mathrm{U}^2}/\mathrm{I}$) and the polarization angle ($\Psi=\frac{1}{2}\arctan\left[ \frac{\mathrm{U}}{\mathrm{Q}} \right]$). The former measures the net fraction of the incoming radiation that is linearly polarized, while the latter is the direction of the linear polarization electric vector on the sky. In blazars, the polarization degree of synchrotron radiation ($\Pi_\mathrm{Sy}$) ranges from zero to as high as $\sim$50\% (e.g., \citealt{Mead1990_BLLs_high_PD}) and generally indicates how ordered the magnetic field lines are in the emission region. Both electron and proton synchrotron radiation are expected to yield a high degree of polarization if the magnetic field is not very disordered. One way to distinguish them is by measuring the variability of $\Pi_\mathrm{Sy}$, since that of the proton synchrotron is expected to be more stable due to the longer proton cooling times compared to electrons and positrons (\citealt{Zhang2016_hadronic_pol_stable}).

On the other hand, while Compton scattering does not inherently lead to polarized radiation, it can do so depending on the polarization of the seed photons and the anisotropy of the upscattering particles. In SSC, since the seed photons are polarized, $\Pi_\mathrm{SSC}$ is expected to be non-zero but still much smaller than $\Pi_\mathrm{Sy}$ ($\Pi_\mathrm{SSC}/\Pi_\mathrm{Sy} \sim 0.3$; e.g., \citealt{bonometto1973,poutanen1994,Krawczynski2012,Peirson2019_PDssc_to_PDsy}). In EC, while the seed photons are thermal and unpolarized, the upscattering electrons may be relatively cold and exhibit extreme anisotropy with respect to the seed photons, which could result in a sizable value of $\Pi_\mathrm{EC}$ (e.g., \citealt{begelman1987}). Alternatively, if such anisotropy does not exist, then $\Pi_\mathrm{EC}$ is expected to be very small ($<1\%$; e.g., \citealt{bonometto1970,Krawczynski2012}).

Such contrasts between the polarization properties of the various emission models mean that contemporaneous MWL polarization measurements can serve as powerful diagnostic tools for constraining emission and acceleration models, as well as the magnetic field strength and morphology in blazar jets, especially when probing higher-energy bands, such as X-rays and $\gamma$-rays (e.g., \citealt{zhang2013,Zhang2019}). Notably, in HSP blazars the peak of the low-energy (synchrotron) component falls close to the X-ray regime; hence, HSP sources are expected to show a high degree of polarization in the X-ray band. This expectation was confirmed with the advent of the Imaging X-ray Polarimetry Explorer (IXPE, \citealt{Weisskopf2022_ixpe_technical}), which has opened a new window of polarimetric measurements to the X-ray universe (2\textendash8~keV; e.g., \citealt{Ehlert2022CenA}); the highest detected $\Pi_\mathrm{X}$ thus far is 31\%, in the HSP blazar PKS~2155$-$304 (\citealt{Kouch2024_pks2155}). Altogether, MWL campaigns of several HSP sources conducted contemporaneously by the IXPE collaboration have revealed a general pattern: while there is a lack of correlation between the MWL polarization properties, the polarization degree in the X-ray band is on average greater than that in the optical band, which is itself (on average) greater than that in the radio band (i.e., $\Pi_\mathrm{X} > \Pi_\mathrm{O} \gtrsim \Pi_\mathrm{R}$). This favors acceleration at a shock front as the most likely scenario of particle energization in the jets of HSP sources, with energy stratification in the region beyond the front where the particles lose energy to radiation and adiabatic expansion (e.g., \citealt{Marscher1985_shock_wave,Liodakis2022,diGesu2022_ixpe_mrk421,Middei2023_ixpe_pg1553,diGesu2023_ixpe_EVPA_rot,Ehlert2023_ixpe_1es0229,Kim2024_ixpe_mrk421,Kouch2024_pks2155,Chen2024_ixpe_mrk501}). On the other hand, in the case of one HSP source, 1ES~1959+650, it exhibited greater variability than could be resolved by IXPE and the aforementioned trend was not detected (\citealt{Errando2024_ixpe_1es1959}).

In LSP and ISP sources, the X-ray regime generally probes the high-energy component or a mix of the low-energy and high-energy components of the blazar emission. Thus, a sizable $\Pi_\mathrm{X}$ (i.e., $\Pi_\mathrm{X} \gtrsim \Pi_\mathrm{O}$) could suggest that hadronic processes play a key role in the high-energy emission of blazars or that the majority of the high-energy emission can be explained by the anisotropic-EC scenario. On the other hand, a low value of $\Pi_\mathrm{X}$ (i.e., $\Pi_\mathrm{X} \sim 0.3 \times \Pi_\mathrm{O}$) in LSPs and ISPs would suggest that their high-energy component is dominated by SSC. Likewise, $\Pi_\mathrm{X} \sim 0$ would strongly favor an EC or SSC dominated emission process where the relativistically moving plasma does not contain relatively cold electrons. The IXPE observations of the blazar BL Lacertae (an LSP-ISP source) in LSP-like states measured $\Pi_\mathrm{X}<12.6\%$ (at 99\% confidence level) and demonstrated that $\Pi_\mathrm{X} < \Pi_\mathrm{O}$ (\citealt{Middei2023_ixpe_bllac}). Meanwhile, during an ISP-like state when the X-ray photons were coming from the low-energy (synchrotron) component, an IXPE observation of BL Lacertae resulted in $\Pi_\mathrm{X} \sim 22\% > \Pi_\mathrm{O}$ in the limited energy range of 2--4 keV (\citealt{Peirson2023_ixpe_bllac}). The IXPE observations of other LSP-ISP sources (three LSPs and one ISP) have only resulted in upper limits of $\Pi_\mathrm{X}<9\%$ to $\Pi_\mathrm{X}<28\%$ (at 99\% confidence levels; \citealt{Marshall2023_ixpe_lsp_isp}). Therefore, while these results are not conclusive enough to completely exclude the presence of some hadronic processes in blazar jets, they strongly disfavor purely hadronic (e.g., pure proton synchrotron) and anisotropic-EC scenarios.


In this paper, we present the results of a MWL campaign of the IXPE collaboration on the LSP blazar \blz\ during an optical-X-ray flaring episode. \blz\ has J2000 sky coordinates of $\mathrm{RA}=149.697083^{\circ}$ and $\mathrm{Dec}=65.565278^{\circ}$ at a redshift of $0.3694 \pm 0.0011$ (\citealt{Lawrence1986_original_z_estimate,becerragonzalez2021}). It is known to frequently exhibit high polarization and flaring behavior in the optical band. It was identified among the prime targets for IXPE to distinguish between leptonic and hadronic scenarios (\citealt{Liodakis2019_detection_probability}). \blz\ is generally classified as a BLL due to its rather weak emission lines (\citealt{Ghisellini2011}). Nevertheless, the features of its SED somewhat resemble those of an FSRQ, which suggests that \blz\ may be a transitional object between the two classes (\citealt{becerragonzalez2021}). \blz\ is a known emitter of GeV and VHE $\gamma$-rays (\citealt{Mukherjee1995,Krauss2014ATel,magic2018}). Photometric observations of this source have found long-term variability in both the radio and optical bands (with the former lagging behind the latter by $\sim$3 weeks), which was interpreted to arise from dynamical rotation of the jet that gradually alters the beaming angle with respect to our line of sight (e.g., \citealt{Raiteri1999_first_SED_modelling,Raiteri2021_WEBT_opt_variability_study,Raiteri2023_WEBT_opt_variability_study}). Timescale analyses by \cite{Sharma2024_possible_disc_jet_coupling} of the 15-year-long \textit{Fermi}-LAT light curves of \blz\ have led to the speculation that the data indicate disc-jet coupling. Furthermore, both photometric and polarimetric observations of \blz\ have shown rapid variability in both the radio and optical bands (e.g., \citealt{Gabuzda2000_radio_pol_variability,Marchili2012_change_in_characteristic_of_IDV,Bachev2015_fast_opt_var_during_2015_flare,Raiteri2021_WEBT_opt_variability_study,Pandey2023_extremely_fast_opt_var,Raiteri2023_WEBT_opt_variability_study}). 

\cite{Morozova2014_fast_flare_of_2011_knot_ejection} observed \blz\ as part of a MWL campaign in 2011. They performed optical photometry and polarimetry measurements, as well as Very Long Baseline Array (VLBA) imaging of the source, and utilized \textit{Fermi} $\gamma$-ray flux density measurements in their study. Between January and March, the optical brightness of \blz\ increased by $\sim$3~mag. Between  March and April 2011, they observed several optical-$\gamma$-ray flares, accompanied by smooth rotations of the optical polarization angle ($\Psi_\mathrm{O}$). On the night of March 9, \blz\ exhibited an extremely fast optical-$\gamma$-ray flare (its optical brightness increased by $\sim$0.7~mag within 7~h); this occurred along with a smooth $300^\circ$ rotation in $\Psi_\mathrm{O}$. This was likely associated with the ejection of an apparently superluminal ($\sim$19$c$) knot from the compact ``core'' of \blz. Furthermore, they analyzed other optical flares between 2008 and 2012, finding similar connections between optical variability and the ejection of superluminal knots from the core. Notably, they found that \blz\ generally shows pronounced optical variability (both in flux and polarization properties). Interestingly, they also determined that \blz\ tends to eject around three superluminal knots per year, similar to the rate of optical flares.

The \cite{magic2018} conducted a MWL campaign on \blz\ in February 2015, when it was experiencing a major flare from radio to VHE $\gamma$-rays. They interpreted the flare to be caused by a relativistic plasma blob (detected in high-resolution 43~GHz maps of the compact core) propagating in a helical magnetic field and crossing a standing shock. This event was accompanied by an optical polarization angle rotation of $\sim$$100^{\circ}$. They modeled the high-energy emission by invoking EC models where the IR seed photons originate from a luminous dusty torus. Similarly, \cite{Deng2022_BLR_contributing_to_EC} showed that EC models could still explain the observed SED of the source even when using far more energetic seed photons, possibly coming from a putative broad-line region (\citealt{becerragonzalez2021}) residing closer to the central engine than the dusty torus. Other studies invoked a combination of SSC and EC models. For example, the significant hardening of the spectrum at GeV energies and the softening at keV energies during this major flaring episode in 2015 were interpreted by \cite{Tanaka2016_fermi_SED_analysis_during_2015_flare} as favoring SSC models, although they used a one-zone, dusty-torus-driven EC model to explain the GeV-TeV portion of the SED.

\cite{Raiteri2023_WEBT_opt_variability_study} conducted a high-cadence photometric and polarimetric monitoring campaign on \blz\ in the optical band during another major flaring episode (April-May 2022). They detected an extremely short variability timescale of 17~minutes, and used it to estimate the size of the emission region to be on the order of $10^{-4}$~pc. They linked such fast, bluer-when-brighter variability patterns to SSC processes present in the jet of the source. Additionally, they found that the variations of the polarization properties in the optical band were more extreme than the radio band, but the average optical polarization angle ($\Psi_\mathrm{O}$) and radio polarization angle ($\Psi_\mathrm{R}$) remained similar. During this flaring period, on April 13, 2022 , \cite{Pandey2023_extremely_fast_opt_var} detected a $120^{\circ}$ polarization angle rotation within a 3~hour period in the optical band. They also measured a change as large as $\sim$10\% in the optical polarization degree of the source within one day. Additionally, they observed that, from mid-March to the end of April 2022, \blz\ reached minimum and maximum values of $\Pi_\mathrm{O}$ of 3\% and 39\%, respectively. In this period, the range of $\Psi_\mathrm{O}$ was $11^\circ$ to $169^\circ$. Notably, no correlation between the optical flux density and $\Pi_\mathrm{O}$ variations was found (\citealt{Raiteri2021_WEBT_opt_variability_study, Pandey2023_extremely_fast_opt_var}).

Milliarcsecond-scale (mas) very-long-baseline interferometry (VLBI) maps of \blz\ in the 5 and 8~GHz bands in the 1990s revealed a bright, compact ``core,'' plus a jet pointing west-northwest, bending with distance to point almost directly westward (i.e., $-90^{\circ} \lesssim \mathrm{\Psi_{jet,5}} \lesssim -70^{\circ}$)\footnote{Position angles are measured from north through east.} 3-5 mas from the core \citep{Gabuzda2000_radio_pol_variability}. 15~GHz VLBI maps of \blz, as part of the Monitoring Of Jets in Active galactic nuclei with VLBA Experiments (MOJAVE\footnote{\href{https://www.cv.nrao.edu/MOJAVE/sourcepages/0954+658.shtml}{https://www.cv.nrao.edu/MOJAVE/sourcepages/0954+658.shtml}}) program between the years 1995 and 2013, revealed the jet to point in the general northwest direction on mas scales ($-60^{\circ} \lesssim \mathrm{\Psi_{jet,15}} \lesssim -30^{\circ}$; \citealt{Hovatta2014_mojave, Lister2018_mojave}). In 43~GHz VLBA images on sub-mas scales between 2010 and 2018, the jet was in the north-northwest direction ($ -30^{\circ} \lesssim \mathrm{\Psi_{jet,43}} \lesssim -10^{\circ}$; \citealt{Jorstad2017_vlbi_maps_2010, magic2018, Weaver2022}). The most recent 43~GHz VLBA maps of the source\footnote{\href{https://www.bu.edu/blazars/VLBA_GLAST/0954.html}{https://www.bu.edu/blazars/VLBA\_GLAST/0954.html}} suggest that the parsec-scale jet of \blz\ still pointed toward the general north-northwest direction throughout 2023 (for more details, see Sects. \ref{sec_mwl_vlbi_maps}).

We describe the X-ray and the accompanying contemporaneous MWL observations of \blz\ in Sects. \ref{sec_x_obs} and \ref{sec_opt_obs}, respectively. We explain the process used to theoretically model the emission of the source in Sect. \ref{sec_models}. Finally, we summarize the results and discuss them in the context of leptonic and hadronic models in Sect. \ref{sec_discussion}.

\section{X-ray observations and analysis methods} \label{sec_x_obs}

\subsection{IXPE observations} \label{sec_x_obs_ixpe}

\begin{figure}
    \centering
    \includegraphics[width=0.49\textwidth]{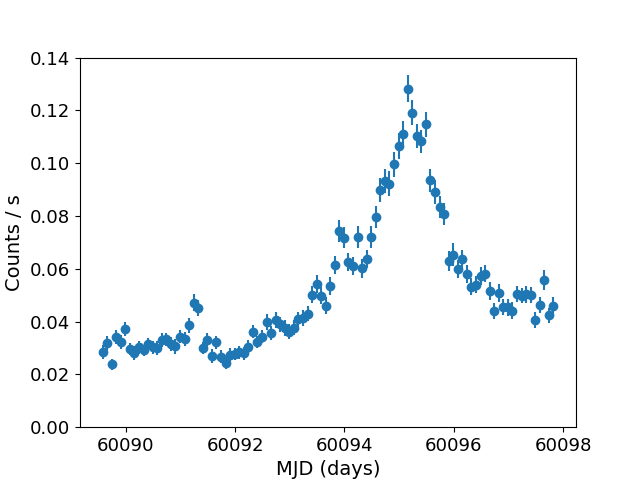}
    \caption{X-ray photon counts from \blz\ as detected by IXPE plotted against time (MJD). IXPE observed the source from MJD 60089 until 60098. The X-ray flux experiences several minor flares and a major one. In this paper, we focus on the major flare which occurs approximately from MJD 60094 to 60096 (peaking at MJD 60095, 2023 May 31).}
    \label{fig_IXPE_flux}
\end{figure}

IXPE observed \blz\ between  May 25, 2023 (MJD 60089) and June 03, 2023 (MJD 60098) with a total exposure time of about 490~ks. We remove the instrumental background following the rejection strategy of \cite{dimarco2023} and we use the \texttt{ixpeobssim} software \citep{baldini2022} to analyse the data. The source events are extracted within a circular region (radius of 1.1~arcmin) centered on the source, while the background events are determined using an annulus with inner and outer radii of 2.0 and 5.0~arcmin, respectively, centered on the source. Figure~\ref{fig_IXPE_flux} shows the light curve of the source in the 2--8 keV range obtained using the \texttt{LC} algorithm of \texttt{ixpeobssim}. The source exhibits a flare between MJD 60094 and 60096 with a peak at MJD 60095 (i.e., 2023 May 31).

We determined the polarization properties (in the 2--8~keV energy regime) using the procedure described in \citet{Kislat2015}, which is implemented in the \texttt{PCUBE} algorithm of \texttt{ixpeobssim}. Additionally, we obtained the 2--8~keV polarization properties using a spectro-polarimetric fit. The two procedures return consistent results within uncertainties. Henceforth, we focus only on the results of the spectro-polarimetric fit. The energy spectrum was obtained from the \texttt{PHA1} algorithm of \texttt{ixpeobsim}, while \texttt{PHA1Q} and \texttt{PHA1U} were used to derive spectra for each of the Stokes parameters. Following \cite{DiMarco2022}, a weighted analysis was employed to improve the sensitivity. For the total energy spectrum (I-spectrum), the bins were merged with \texttt{ftgrppha} of \texttt{ftools} until a signal-to-noise ratio (S/N) of at least 5 is reached. For the spectra of the Stokes parameters (Q- and U-spectra), a binning was chosen to guarantee at least five counts in each bin. The spectro-polarimetric fit is performed via \texttt{Xspec}. To account for the Galactic absorption, we used the Tuebingen-Boulder model (\texttt{tbabs} in \texttt{Xspec}) with an $n_H$ column density of $4.68 \times 10^{20}$ cm$^{-2}$ \citep{NhAbsorption}. We fit the data with a power law model (\texttt{powerlaw}) multiplied by an energy-independent polarization (\texttt{polconst}).

We did not detect X-ray polarization ($\Pi_\mathrm{X}$) of \blz\ at a $3\sigma$ confidence level. When taking the entire IXPE observation window into account, we obtained an upper limit (99.73\% confidence level) of $\Pi_\mathrm{X} < 14\%$. In an attempt to improve the constraints on $\Pi_\mathrm{X}$, we reduced the number of free parameters involved in the spectro-polarimetric fit by assuming a number of physical scenarios: we fix the X-ray polarization angle ($\Psi_\mathrm{X}$) to be either parallel or perpendicular to the jet position angle ($\Psi_\mathrm{jet,43}=-42^\circ \pm 7^\circ$, see Sect. \ref{sec_mwl_vlbi_maps}) or to the measured optical polarization angle ($\Psi_\mathrm{O}=66.5^\circ$, see Sect. \ref{sec_mwl_vlbi_maps}). Additionally, we consider three time intervals: the full IXPE exposure (from MJD 60089 to 60098); the interval in which the optical polarization angle appears to stay constant (MJD 60092 to 60095, see Sect. \ref{sec_opt_obs}); and the period during which the X-ray flux peaks (as detected by IXPE, from MJD 60094 to 60096; see Fig.~\ref{fig_IXPE_flux}). The results of these tests are presented in Table \ref{tab_IXPE_results}. A scenario under which $\Psi_\mathrm{X}$ is roughly perpendicular to the constant value of $\Psi_\mathrm{O}$ yields the strongest constraints on the X-ray polarization degree, $\Pi_\mathrm{X}<8.4\%$ (at 99.73\% confidence level). Notably, when $\Psi_\mathrm{X}$ is perpendicular to the jet position angle, we obtained $\Pi_\mathrm{X}<8.8\%$ (at a 99.73\% confidence level). We emphasize that fixing other values of $\Psi_\mathrm{X}$ typically has a minimal effect on the $\Pi_\mathrm{X}$ upper limits. Generally, we  obtained less constraining upper limits when the shorter time bins are considered.

Since the source exhibits a flare during the IXPE observation, we inspect the data for spectral variability during the observation. All photon spectral index measurements, as calculated with the spectro-polarimetric fit, are consistent with a constant value of $\Gamma=1.69 \pm 0.03$. Similarly, all hardness ratio measurements, defined as $(\phi_{\mathrm{low}} - \phi_{\mathrm{high}}) $/$ (\phi_{\mathrm{low}} + \phi_{\mathrm{high}}$) with $\phi_{\mathrm{low}}$ and $\phi_{\mathrm{high}}$ being the photon count rates in the 2--4 and 4--8~keV intervals, respectively, remain consistent with a constant value of 0.71$\pm$0.01 throughout the IXPE observation. This indicates that the source did not experience any significant spectral variations. Therefore, we checked for variability in polarization by splitting the IXPE data into both equal and variable time intervals. To obtain the polarization properties in this test, we used both the spectro-polarimetric fit and the \texttt{PCUBE} algorithm. No X-ray polarization was detected in any of the selected time intervals via either of the two methods. Lastly, to check for energy dependence, we divided the IXPE data into two ranges: 2--4 and 4--8~keV. We then applied the spectro-polarimetric fit to both separately, using an energy-independent polarization model (\texttt{polconst} in \texttt{Xspec}). We did not find any significant polarization in either of the two energy bins, as $\Pi_\mathrm{X}$ upper limits (at the 3$\sigma$ level) are $>$20\% in all of the tests.

\begin{table}
\setlength{\tabcolsep}{1.9pt}
\centering
\caption{Upper limits (99\% and 99.73\% confidence levels) to the X-ray polarization angle ($\Psi_\mathrm{X}$) and degree ($\Pi_\mathrm{X}$) as calculated via the spectro-polarimetric fit for different alignment scenarios. \label{tab_IXPE_results}}
   \begin{tabular}{c | l l c c}
   \hline\hline
   T & $\Psi_\mathrm{X}$ alignment & Fixed $\Psi_\mathrm{X}$ & $\Pi_\mathrm{X,99\%}$ & $\Pi_\mathrm{X,99.73\%}$ \\
    (1) & (2) & (3) & (4) & (5) \\
   \hline
    \multirow{5}{*}{\rotatebox[origin=c]{90}{Full}} & $\Psi_\mathrm{X} \parallel \Psi_\mathrm{jet,43}$ & $-42^\circ \equiv 138^\circ$ & $<9.8$\% & $<11$\% \\
     & $\Psi_\mathrm{X} \perp \Psi_\mathrm{jet,43}$ & $48^\circ \equiv -132^\circ$ & $<7.4$\% & $<8.8$\% \\
     & -- & (not fixed) & $<13$\% & $<14$\% \\
     & $\Psi_\mathrm{X} \parallel \Psi_\mathrm{O}$ & $66.5^\circ \equiv -113.5^\circ$ & $<10$\% & $<11$\% \\
     & $\Psi_\mathrm{X} \perp \Psi_\mathrm{O}$ & $156.5^\circ \equiv -23.5$ & $<7.0$\% & $<8.4$\% \\
    \hline
    \multirow{3}{*}{\rotatebox[origin=c]{90}{Const.}} & -- & (not fixed) & $<20$\% & $<22$\% \\
     & $\Psi_\mathrm{X} \parallel \Psi_\mathrm{O}$ & $66.5^\circ \equiv -113.5^\circ$ & $<8.1$\% & $<9.9$\% \\
     & $\Psi_\mathrm{X} \perp \Psi_\mathrm{O}$ & $156.5^\circ \equiv -23.5$ & $<17$\% & $<19$\% \\
     \hline
    \multirow{3}{*}{\rotatebox[origin=c]{90}{Peak}} & -- & (not fixed) & $<13$\% & $<15$\% \\
     & $\Psi_\mathrm{X} \parallel \Psi_\mathrm{O}$ & $66.5^\circ \equiv -113.5^\circ$ & $<12$\% & $<14$\% \\
     & $\Psi_\mathrm{X} \perp \Psi_\mathrm{O}$ & $156.5^\circ \equiv -23.5$ & $<11$\% & $<13$\% \\
    \hline
        \end{tabular}
    \tablefoot{Columns: (1) time interval used in the spectro-polarimetric analysis; (2) alignment used to fix the value of the X-ray polarization angle ($\Psi_\mathrm{X}$); (3) fixed value of $\Psi_\mathrm{X}$; (4) upper limit to X-ray polarization degree at 99\% confidence level; (5) upper limit to X-ray polarization degree at 99.73\% confidence level (i.e., Gaussian $3\sigma$). ``Full'' refers to the entire IXPE observing period (MJD 60089-60098); ``Const.'' refers to the constant $\Psi_\mathrm{O}$ period (MJD 60092-60095), and ``Peak'' refers to the period in which the IXPE flux peaked (MJD 60094-60096; see Fig.~\ref{fig_IXPE_flux}). In column (3), the symbol "$\equiv$" is used to represent the inherent $\pm 180^\circ$ ambiguity in the polarization angles.}
\end{table}

\subsection{\textit{XMM-Newton}} \label{sec_x_obs_xmm}

The \textit{XMM-Newton} space observatory \citep{Jansen2001} targeted \blz\ from 2023 May 20 (22:41:00~UT) to May 21 (01:44:53~UT) for a net exposure time of 7.7~ks (Obs. ID 0920901901). The standard routines of the \texttt{SAS} (Science Analysis System, Version 21.0.0) and the latest calibration files are used to calibrate and extract the third-level science products. The background is determined from a circular region with a 40~arcsec radius centered on a source-free area of the CCDs. Then, a circular region with a radius of 22~arcsec is adopted to extract the spectrum of \blz. This radius is selected based on an iterative process aimed at maximizing the signal-to-noise ratio in the 0.3 to 10~keV regime (e.g., \citealt{Piconcelli2004}). We subsequently binned the obtained spectrum to have at least 30 counts per bin, while making sure to avoid oversampling the instrumental energy resolution by a factor larger than 3.

We modeled the European Photon Imaging Camera pn-CCD (EPIC-pn; \citealt{Struder2001_xmm_EPIC}) spectrum with an absorbed, pegged power law (\texttt{tbabs*pegpwrlw}). Additionally, since it is known that there is a strong correlation between the photon index and the absorption parameter, we fixed the value of the Galactic hydrogen absorption to $N_\mathrm{H,Gal}=4.68\times10^{20}\,\mathrm{cm}^{-2}$ \citep{HI4PI}. We calculated the C-statistics (\citealt{cash1979}) of the fit and, to assess the goodness of the fit, we additionally computed the theoretical C-statistics as well as its variance, following \cite{Kaastra2017}. The best-fit result is given in Appendix \ref{appendix_Swift_XMM_spectra}.

\subsection{\textit{Swift}-XRT} \label{sec_x_obs_swift}

During the IXPE observation, we also observed \blz\ using the X-ray Telescope (XRT) on board the space-borne \textit{Neil Gehrels Swift} Observatory (\textit{Swift}-XRT; see \citealt{Gehrels2004_swift}) in photon-counting mode. Data taken with \textit{Swift}-XRT is reduced and extracted following the standard procedures, using \texttt{xrtpipeline} (Version 0.13.7) and \texttt{HEASOFT} (Version 6.30). We calculated the photon flux and spectrum from a circular region centered on the source with a radius of 35~arcsec. The background events were extracted using an annulus centered on the source with inner and outer radii of 50 and 150~arcsec, respectively. The spectra were binned such that each bin contained at least one photon and they were individually fit using C-statistics \citep{cash1979}. To include the foreground absorption of the interstellar medium, we use \texttt{vern} cross sections \citep{vern1996} and \texttt{wilm} abundances \citep{wilms2000_tbabs}. We fit the data with the same model as described for the \textit{XMM-Newton} spectrum. The best-fit results are given in Appendix \ref{appendix_Swift_XMM_spectra}.

\section{Radio, optical, and $\gamma$-ray observations} \label{sec_opt_obs}
\subsection{Optical and radio data} \label{sec_opticalradio}
Contemporaneous with the IXPE observations, \blz\ was also observed in the radio and optical bands using the Alhambra Faint Object Spectrograph and Camera (ALFOSC) at the Nordic Optical Telescope (NOT), the Effelsberg 100-m telescope \citep{Kraus2003}, the LX-200 telescope operated by St.\ Petersburg State University, RoboPol at the Skinakas Observatory \citep{Ramaprakash2019_Skinakas_RoboPol}, the T90 telescope at the Sierra Nevada Observatory (SNO), and the Submillimeter Array (SMA, \citealp{Ho2004}). Details on the data analysis and observing strategy for these MWL observations can be found in \cite{Liodakis2022,Middei2023_ixpe_bllac,diGesu2023_ixpe_EVPA_rot,Kouch2024_pks2155}.

\begin{figure*}
    \centering
    \includegraphics[width=\textwidth]{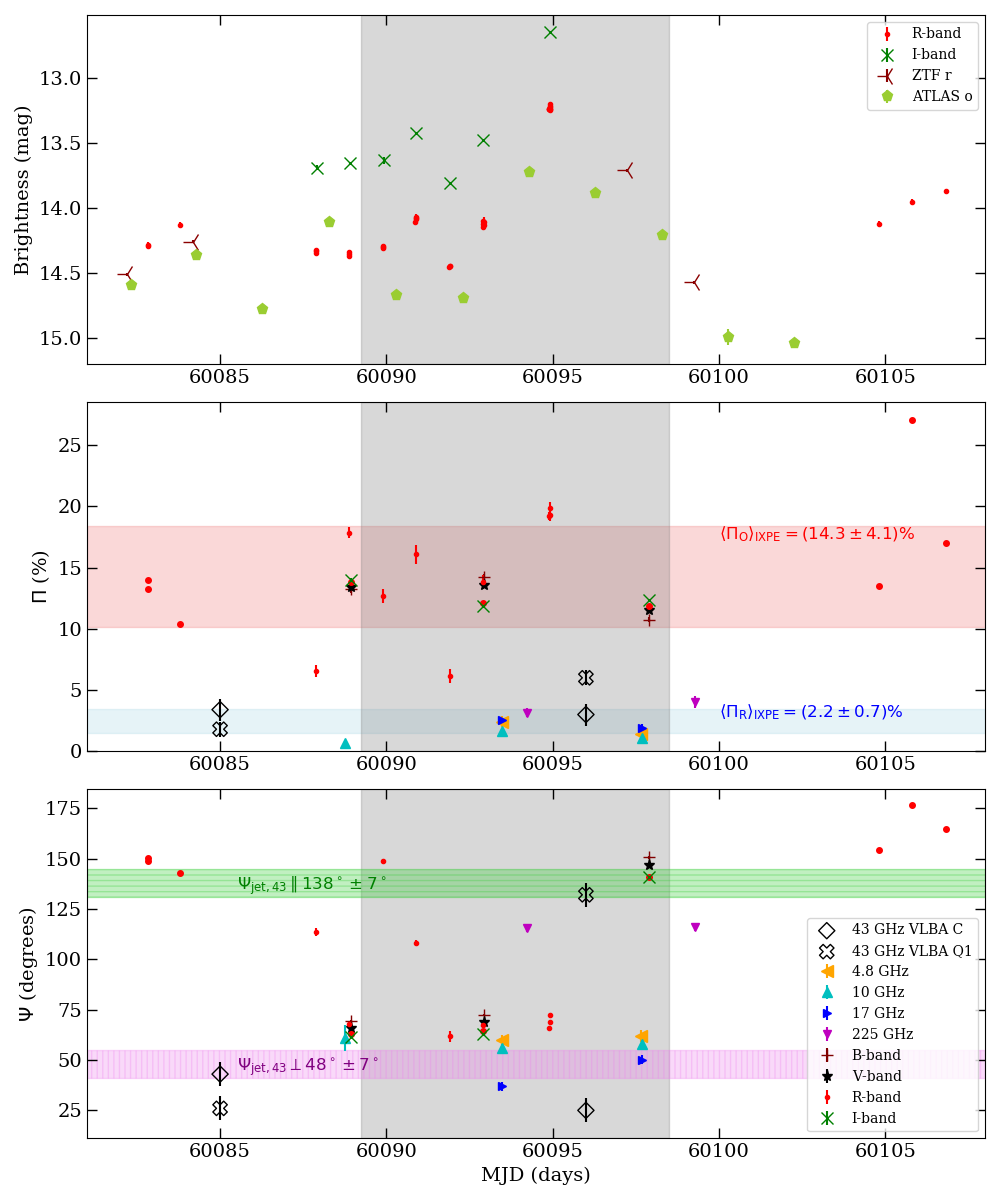}
    \caption{Contemporaneous radio and optical observations of \blz. From top to bottom, the panels show optical brightness in magnitudes, polarization degree ($\Pi$) in \%, and polarization angle ($\Psi$) in degrees. The grey shaded area (vertical) marks the duration of the IXPE observation (MJD 60089-60098). In the middle panel, the red and blue shaded areas (horizontal) are centered on the average $\Pi_\mathrm{O}$ and $\Pi_\mathrm{R}$ within the IXPE observation time window, respectively; the thickness of these red and blue horizontally shaded areas corresponds to the standard deviation of $\Pi_\mathrm{O}$ and $\Pi_\mathrm{R}$ within the IXPE observation time window, respectively. In the bottom panel, we indicate the parallel and perpendicular alignment directions of the jet position angle (after shifting $\Psi_\mathrm{jet,43}=-42^\circ \pm 7^\circ$ by $180^\circ$ to account for the inherent $180^\circ$ ambiguity in polarization angle estimates; see Sect. \ref{sec_mwl_vlbi_maps}) using the green and purple horizontally shaded areas, respectively. The 43~GHz VLBA C and Q1 data points refer to the polarization degree and angle of the two parsec-scale components detected at 43~GHz (see Sect. \ref{sec_mwl_vlbi_maps}).}
    \label{fig_all}
\end{figure*}

The Effelsberg observations were obtained as part of the Monitoring the Stokes Q, U, I and V Emission of AGN jets in Radio (QUIVER) program (\citealt{Myserlis2018}), and performed at 4.8, 10, and 17~GHz. The SMA observations were obtained at 225.5~GHz within the SMA Monitoring of AGNs with POLarization (SMAPOL) program (\citealt{Marrone2008, Myserlis2024_poster}). In the B, V, R, and I optical bands ($\lambda_\mathrm{eff}$ of 445, 551, 658, and 806~nm, respectively), NOT provided polarimetric observations (for analysis details see, e.g., \citealt{Hovatta2016,Nilsson2018}). Moreover, LX-200, RoboPol, and SNO provided R-band polarimetric observations. Additionally, SNO provided photometry in R-band, and LX-200 in both R- and I-bands. We also have optical photometry data from two all-sky surveys via performing forced-photometry at the coordinate of \blz: r-band data ($\lambda_\mathrm{eff}=637$~nm) from the Zwicky Transient Facility (ZTF\footnote{\href{https://irsa.ipac.caltech.edu/Missions/ztf.html}{https://irsa.ipac.caltech.edu/Missions/ztf.html}}; \citealt{bellm2019_ZTF}) and o-band ($\lambda_\mathrm{eff}=578$~nm) data from the Asteroid Terrestrial-impact Last Alert System (ATLAS\footnote{\href{https://fallingstar-data.com/}{https://fallingstar-data.com/}}; \citealt{tonry2018_ATLAS}) all-sky survey. In Appendix \ref{appendix_mwl_data}, the optical photometric (in magnitudes) and polarimetric measurements of \blz\ are given in Tables \ref{tab_opt_data1} and \ref{tab_opt_data2}, while Table \ref{tab_radio_results} summarizes the polarimetric observations of the source made in the radio bands. The time dependence of the contemporaneous MWL polarization measurements of \blz\ is shown in Fig.~\ref{fig_all}.

For about one month prior to the IXPE observations of \blz, the optical brightness was stable at $\sim$14.5~mag, as observed by T90 (see Table \ref{tab_opt_data1}). However, during the IXPE observation (MJD 60089-60098), \blz\ exhibited a flare in all of the observed optical bands, with its R-band brightness rising from $\sim$14.5~mag to $\sim$13.2~mag from MJD 60092 to 60095. During the entire IXPE pointing, the optical polarization degree ($\Pi_\mathrm{O}$) and angle ($\Psi_\mathrm{O}$) were significantly variable, which is similar to the past behavior of \blz\ (e.g., \citealt{Raiteri2023_WEBT_opt_variability_study, Pandey2023_extremely_fast_opt_var}). For example, during this 1~mag optical flare, $\Pi_\mathrm{O}$ rose from $\sim$6\% to  $\sim$20\%. During the entire IXPE pointing, $\Pi_\mathrm{O}$ fluctuated about an average of $14.3\%$ with a standard deviation of $4.1\%$. Meanwhile, $\Psi_\mathrm{O}$ varied from $62^\circ$ just before the start of the IXPE observation, to $149^\circ$ ($\Delta\Psi_\mathrm{O}\approx85^\circ$) within one day. Right after this fast rotation, $\Psi_\mathrm{O}$ returned to $66.5^\circ$ ($\Delta\Psi_\mathrm{O}\approx87^\circ$ within 2~d) and remained constant for about 3~d (coinciding with the 1~mag flare). By the end of the IXPE observation (MJD 60098), $\Psi_\mathrm{O}$ returned to $\sim$$140^\circ$ and, a few days later, to $\sim$$170^\circ$. We note that, five days prior to the IXPE pointing, $\Psi_\mathrm{O}$ was $\sim$$150^\circ$. Since the 43~GHz parsec-scale jet of \blz\ points toward the northwest direction ($\Psi_\mathrm{jet,43} = -42^\circ \pm 7^\circ$; see Sect. \ref{sec_mwl_vlbi_maps}), the temporal behavior of the optical polarization angle can be summarized roughly in the following manner (also see Fig.~\ref{fig_all}):
\begin{itemize}
    \item MJD 60084: $\Psi_\mathrm{O} \parallel \Psi_\mathrm{jet,43}$ (followed by a rotation over 5~d);
    \item MJD 60089: $\Psi_\mathrm{O} \perp \Psi_\mathrm{jet,43}$ (followed by a rotation in 1~d);
    \item MJD 60090: $\Psi_\mathrm{O} \parallel \Psi_\mathrm{jet,43}$ (followed by a rotation over 2~d);
    \item MJD 60092: $\Psi_\mathrm{O} \perp \Psi_\mathrm{jet,43}$ (stayed constant for 3~d);
    \item MJD 60095: $\Psi_\mathrm{O} \perp \Psi_\mathrm{jet,43}$ (followed by a rotation over 3~d);
    \item MJD 60098: $\Psi_\mathrm{O} \parallel \Psi_\mathrm{jet,43}$ (stayed constant for at least 9~d).
\end{itemize}

At the low radio frequencies (4.8, 10, and 17~GHz), the degree of polarization of \blz\ during and near the IXPE observation window was  low ($1\%<\Pi_\mathrm{\le 17}<3\%$), with a polarization angle $37^\circ<\Psi_\mathrm{\le 17}<60^\circ$ (i.e., $\Psi_\mathrm{\le 17} \perp \Psi_\mathrm{jet,43}$, roughly). At 225.5~GHz, the polarization degree was slightly higher ($3\% <\Pi_\mathrm{225}< 4\%$), with a polarization angle $\Psi_\mathrm{225}=116^\circ \pm 2^\circ$. Over the course of this MWL campaign, we do not find any significant variability in the polarimetric properties of \blz\ in the radio bands. Hence, during the 1~mag optical flare mentioned above (MJD 60092-60095), $\Pi_\mathrm{R}$ remained stable below 3\%, and $\Psi_\mathrm{R}$ was essentially parallel to the constant $\Psi_\mathrm{O}$ of $66.5^\circ \equiv -113.5^\circ$ and perpendicular to the jet position angle. The lack of variability in the radio polarimetric properties as compared to those of the optical band is in line with the previous observations of \blz\ (\citealt{Raiteri2023_WEBT_opt_variability_study}).

\subsection{Fermi $\gamma$-ray data} \label{sec_gamma}

Using data from the Large Area Telescope (LAT) instrument on board the \textit{Fermi} Gamma-ray Space Telescope \citep{Abdollahi2020}, we obtain $\gamma$-ray flux measurements of \blz\ contemporaneous with the IXPE observation. These fluxes are incorporated into the modeling of the spectral energy distribution of \blz\ (see Sect. \ref{sec_models}). To obtain the $\gamma$-ray fluxes, we employ standard analysis tools (Pass8 \texttt{P8R3\_source}, \textsc{Fermitools} v.1.2.23) to select events in the 100~MeV to 300~GeV energy range. We apply the recommended zenith angle cut (90$^{\circ}$) to avoid contamination from Earth's limb and include the diffuse and isotropic background components\footnote{\url{https://fermi.gsfc.nasa.gov/ssc/data/access/lat/BackgroundModels.html}} to extract the source spectrum via a binned likelihood analysis, assuming a power-law spectrum in each bin. During this analysis, we simultaneously fit the spectrum and normalization of all the sources within the region of interest (ROI), while all sources within 10$^{\circ}$ from the ROI were fixed to the 4FGL-DR2 catalogue parameters \citep{Ballet2020}. In Table \ref{tab_fermi_nuFnu} we present the spectral flux density values obtained in the \textit{Fermi} $\gamma$-ray band.

\begin{figure*}
    \centering
    \includegraphics[width=18cm]{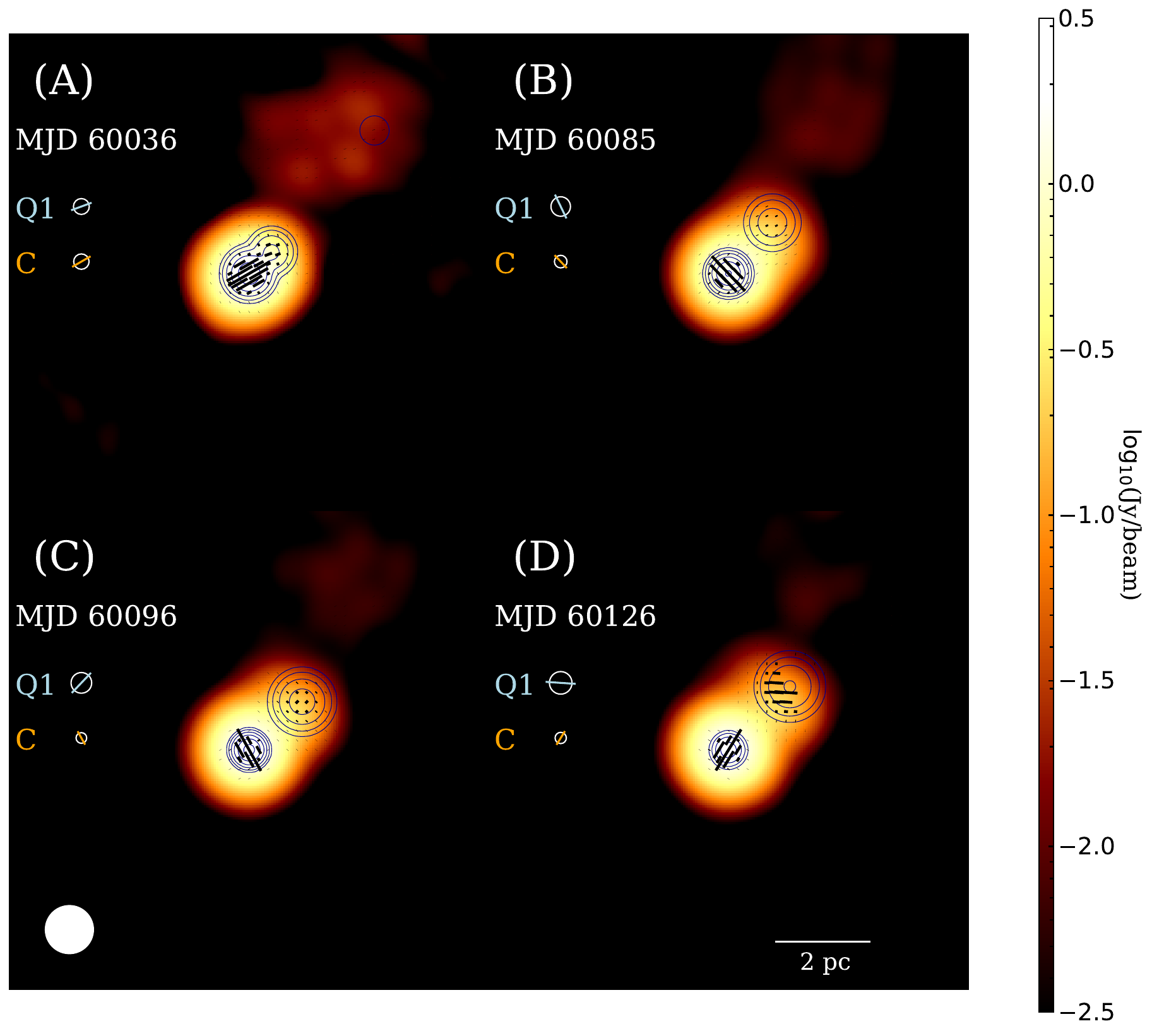}
    \caption{43~GHz total intensity (Stokes I) VLBA images of \blz. Panels (A), (B), (C), and (D) show the parsec-scale maps of the compact core of the source from data obtained on 2023 April 2 (MJD 60036; about two months before the IXPE observation), 2023 May 21 (MJD 60085; five days before the IXPE pointing), 2023 June 1 (MJD 60096; during the IXPE pointing and in the vicinity of the peak of the optical-X-ray flare), and 2023 July 1 (MJD 60126; around one month after the IXPE pointing), respectively. The bottom-left white circle is has a radius of 0.2~mas and is the FWHM of the common convolving beam. The circular contours are Stokes I, between $10^{-6}$ and $10^{-3}$~Jy in log steps. The general direction of the jet is consistently northwest ($ -42^{\circ} \pm6^{\circ}$). In panel (A), a component (Q1) north-northwest of the compact core (C) appears to separate from the core after being ``ejected.'' The polarization angles of both C and Q1 are $\sim$$115^\circ$. In panel (B), the polarization angles of both C and Q1 appear to have rotated to $\sim$$35^\circ$. In panel (C), the polarization angle of C remains comparable to its value in (B), but that of Q1 has rotated by $\sim$$100^\circ$ compared to its value in (B). In panel (D), the polarization angle of C is $132^\circ$, while that of Q1 is $86^\circ$. In each panel, the angular size and polarization angle of the two components are represented by the circular symbols on the left (orange is used for the core component C and cyan is used for the moving component Q1): the radius of the circle symbolizes the angular size of each component (proportional to the FWHM of the intensity), while the orientation of the line through the circle visualizes the direction of the polarization angle.}
    \label{fig_43GHz_VLBA}
\end{figure*}

\subsection{VLBI images of \blz\ near the IXPE observation} \label{sec_mwl_vlbi_maps}

As part of the Blazars Entering the Astrophysical Multi-Messenger Era monitoring program (BEAM-ME, successor to the VLBA-BU-BLAZAR program, which ran from 2007 to 2020), \blz\ was observed by the VLBA at 43~GHz during and near the IXPE observation window. Figure~\ref{fig_43GHz_VLBA} presents four milliarcsec-scale maps of the compact region of the source at a resolution of $\sim$0.1~mas (for more details see \citealt{Jorstad2016_VLBA_BU_BLAZAR,Jorstad2017_vlbi_maps_2010,Weaver2022}). For a redshift of 0.3694 (see Sect. \ref{sec_intro}), a Hubble constant of 70~km~s$^{-1}$~Mpc$^{-1}$, a matter density parameter of $\Omega_\mathrm{m}=0.3$, and assuming a flat Universe, 0.1~mas translates to a projected distance of 0.51~pc. Additionally, the luminosity distance of \blz\ is calculated to be 1980~Mpc.

These 43~GHz VLBA images of \blz\ (Fig.\ \ref{fig_43GHz_VLBA}) reveal the evolution of the compact region of the jet during three months, from 2023 April to the end of June. They show the motion of a prominent component (Q1) to the northwest of the compact core (C) along position angle $-42^\circ \pm 7^\circ$. This angle is calculated using the positions of the C and Q1 components (see Table \ref{table_VLBA}); its uncertainty is derived using x and y positional uncertainties equivalent to 0.2 times the FWHM size of the components. This direction differs by $28^\circ \pm 8^\circ$ from the previously reported \citep[prior to 2018;][]{Weaver2022} jet direction, which was $\Psi_\mathrm{jet,43}=-13.7^\circ \pm 2.1^\circ$. We have reconstructed the VLBA images of the source using a forward-modeling approach, implemented within the \texttt{eht-imaging} software package (\citealt{Chael16}). This entails determining the best-fit solution of the geometrical model, compared with the visibility amplitudes, closure phases, and fractional polarization, in terms of the minimum of an error function (for further details see, e.g., \citealt{Paraschos24a, Paraschos24c}). The modeling parameters for C and Q1 are summarized in Table \ref{table_VLBA}.

At 43~GHz, the total flux of C dominates over that of Q1 by a factor of $\sim$7. In terms of compactness (as measured by the size of the radius of the FWHM total intensity), Q1 is initially comparable in size to C. Over the course of the next three months, Q1 expands from 0.07 to 0.09 mas FWHM. Q1 separated from C with a roughly constant proper motion of $(0.80 \pm 0.07)$~mas~yr$^{-1}$, which translates to a projected apparent speed of $(18 \pm 2)~c$ in the rest frame of the host galaxy. Assuming a constant, linear proper motion, we estimate that Q1 was ejected on MJD 59994$\pm$4. The average polarization degree ($\Pi$) of C is 2.2\%, while that of Q1 is $3.8\%$. We find that $\Pi_\mathrm{Q1}$ peaks at 6\% on MJD 60096, near the time of the optical-X-ray flare. The polarization angle ($\Psi$) of C remains roughly aligned with that of Q1 on MJD 60036 and 60085 at $115^\circ$ and $35^\circ$, respectively (i.e., $\Psi_\mathrm{C,MJD60036} \parallel \Psi_\mathrm{Q1,MJD60036} \sim 115^\circ$ and $\Psi_\mathrm{C,MJD60085} \parallel \Psi_\mathrm{Q1,MJD60085} \sim 35^\circ$). However, on MJD 60096 $\Psi_\mathrm{C}$ remains near its previous value, while $\Psi_\mathrm{Q1}$ rotates to $132^\circ$ (i.e., $\Psi_\mathrm{C,MJD60096} \perp \Psi_\mathrm{Q1,MJD60096}$, roughly). The polarization degree and angle of C and Q1 are plotted as a function of time in Fig.~\ref{fig_all}.

\begin{table}
\caption{Component parameters of the 43~GHz VLBA maps shown in Fig.~\ref{fig_43GHz_VLBA}.}
\label{table_VLBA}
\centering
\begin{tabular}{cccccrr}
 \hline\hline
 ID & MJD & $\mathrm{F}_0$ & $\mathrm{R}_\mathrm{I}$ & [x, y] & $\Pi$ & $\Psi$\\
 (1) & (2) & (3) & (4) & (5) & (6) & (7)\\
 \hline
 C  &  60036  &  2.01 &  0.06 &  [0.00, 0.00]  &  2.2\% &  110$^\circ$\\
 C  &  60085  &  1.66 &  0.05 &  [0.00, 0.00]  &  3.4\% &  43$^\circ$\\
 C  &  60096  &  1.92 &  0.04 &  [0.00, 0.00]  &  3.0\% &  25$^\circ$\\
 C  &  60126  &  1.93 &  0.05 &  [0.00, 0.00]  &  0.3\% &  149$^\circ$\\
 \hline
 Q1 &  60036  &  0.25 &  0.07 &  [$-$0.06, 0.07] &  3.2\% &  120$^\circ$\\
 Q1 &  60085  &  0.27 &  0.08 &  [$-$0.14, 0.17] &  1.8\% &  26$^\circ$\\
 Q1 &  60096  &  0.28 &  0.09 &  [$-$0.17, 0.19] &  6.0\% &  132$^\circ$\\
 Q1 &  60126  &  0.28 &  0.09 &  [$-$0.20, 0.20] &  4.2\% &  86$^\circ$\\
\hline
\end{tabular}
\tablefoot{Column (1) indicates the component ID (C refers to the core and Q1 refers to the component to the northwest of the core), column (2) gives the MJD of each map, column (3) gives the total flux density of the component in Jy, column (4) gives the radius of the full width at half maximum (FWHM) of the total intensity for each component in mas, column (5) gives the x and y coordinates of the components with respect to the core in mas, column (6) gives the polarization degree of the component in \% (the typical errors on $\Pi$ of C and Q1 are $\pm$0.9\% and $\pm$0.6\%, respectively), and column (7) gives the polarization angle of the component in degrees (the typical errors for $\Psi$ are $\pm$6$^\circ$).}
\end{table}

\section{Theoretical models} \label{sec_models}

In this section, we describe the construction of the spectral energy distribution (SED) $\nu \mathrm{F}_\nu$ (frequency multiplied by flux density at that frequency) and MWL spectral polarization distribution (SPD) of \blz, using only data obtained during the IXPE observation (MJD 60089-60098). The SPD represents how the polarization degree $\Pi$ changes with respect to $\nu$. Subsequently, we describe how we model these MWL spectral distributions based on two general scenarios: leptonic (single and multi-zone) and hadronic (proton + pair synchrotron).

To build the MWL SED and SPD, at each of our observed frequencies we determine the median and standard deviation of both the flux density and polarization degree. For the leptonic models, we use the single-zone \citep{Bottcher2013} and multi-zone \citep{Peirson2019_PDssc_to_PDsy} SSC models. We find that we can adequately fit the high-energy portion of the SED without the need for an EC component, as is typical for BLLs and as was previously invoked for \blz\ (e.g., \citealp{Tanaka2016_fermi_SED_analysis_during_2015_flare}; see also Sect. \ref{sec_intro}). The multi-zone model accounts for depolarization and energy stratification but does not include synchrotron self-absorption effects. In this model, a conical jet with a fixed opening angle ($\theta_{\text{op}}$) and constant Lorentz factor ($ \Gamma $) is divided into magnetic field zones, each observed at a unique viewing angle ($\theta_{\text{obs}} $), and thus has a Doppler factor $D = \frac{1}{\Gamma(1 - \beta \cos \theta_{\text{obs}})}$. The single-zone model uses the post-processing and semi-analytical depolarization procedure described in \cite{zhang2013} and \cite{Zhang2024}, respectively. The hadronic model is post-processed with the same single-zone prescription\footnote{Given that the source was in a flaring state, the single-zone approximation for the hadronic emission is valid, even if the protons have an extended spatial distribution \citep{Zhang2024}.} and mainly includes proton-synchrotron radiation. We note that the single-zone models (SSC and hadronic) do not fit the low-energy tail of the observed SED well, because in single-zone models emission from the large-scale jet is not included. Additionally, synchrotron self absorption is not taken into account when making the polarization calculations in any of the models. This limits our polarization estimates in the SPD to $\gtrsim$$10^{12}$~Hz. Similarly, none of the polarization models includes Klein-Nishina effects; therefore, our $\Pi$ calculations above $\sim$$10^{24}$~Hz are not accurate. 

\begin{figure}
\centering
 \resizebox{\hsize}{!}{\includegraphics[width=\textwidth]{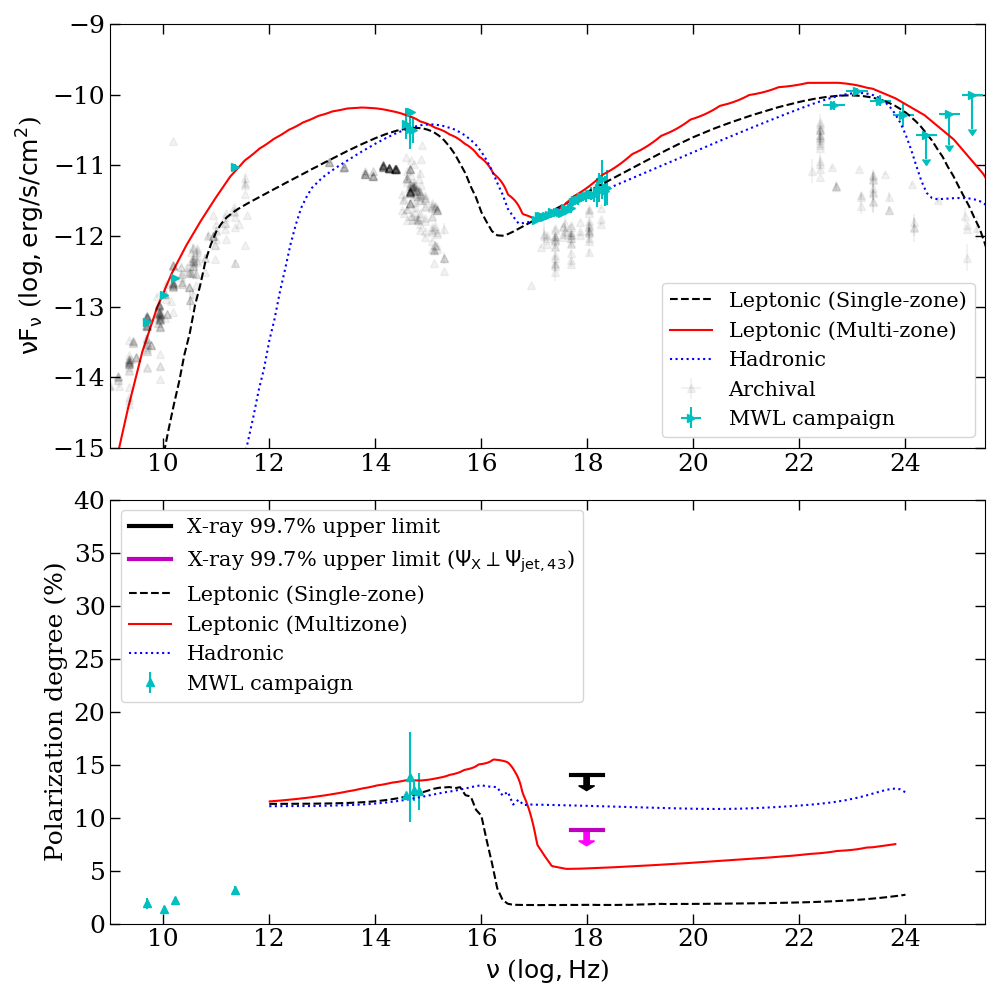}}
 \caption{\textit{Top}: Spectral energy distribution (SED) of \blz. \textit{Bottom}: Spectral polarization distribution (SPD) of \blz. In both panels, the cyan data points show the median and standard deviation values at different frequencies during the IXPE exposure. The archival observations (non-simultaneous) are shown in black for reference. The different models are represented by the solid, dashed, and dotted colored lines, as indicated in the legend. In the bottom panel the black downward arrow indicates the 3$\sigma$ upper limit for the entire exposure, and the magenta arrow the 3$\sigma$ upper limit under the assumption that the X-ray polarization angle is aligned perpendicular to the 43~GHz jet position angle (see Table \ref{tab_IXPE_results}).}
\label{fig_SED_models}
\end{figure}

Table \ref{tab_SEDmodels} summarizes the SED modeling parameters for the different scenarios considered, while Fig.~\ref{fig_SED_models} presents the simultaneous SED and SPD modeling. For leptonic models, we find a drop of $\Pi$ between the low- and high-energy components of the SED, as expected, whereas the hadronic model produces comparable polarization from optical to X-rays. More background information on the polarization behavior in leptonic versus hadronic scenarios is given in Sect. \ref{sec_intro}.

\begin{table*}
\caption{Parameters of the different SED models.}
\label{tab_SEDmodels}
\centering
\begin{tabular}{ll|lll}\hline\hline
 & & SSC (multi-zone) & SSC (single-zone)  & Hadronic \\ \hline
 Bulk Lorentz factor & $\Gamma$                               & 10& 30 & 30  \\ 
 Bulk Doppler factor & $D$                                    & -& 30   & 30  \\ 
 Blob radius & $R\, (\rm{cm})$                                & $2.4\times 10^{15}$& $7\times 10^{16}$  & $ 10^{15}$ \\ 
 Particle escaping time scale & $\tau_{esc}\, (\rm{s})$       &- & $10^{7}$      & $10^{5}$ \\ 
 Magnetic field strength & $B\, (\rm{G})$                     &0.12& 0.028        & 100 \\ \hline
 Electron injection luminosity & $L_e\, (\rm{erg/s})$         & -& $2.6\times10^{46}$  & $1.3\times10^{44}$ \\ 
 Electron lower spectral cutoff & $\gamma_{e,min}$            & 1& $1$  & $3\times10^2$ \\ 
 Electron higher spectral cutoff & $\gamma_{e,max}$           &$1.2\times10^4$ & $2\times 10^4$    & $10^3$ \\ 
 Electron spectral index & $p_e$                              &1.7 & 2.2       & 2.37 \\ \hline
 Proton injection luminosity & $L_p\, (\rm{erg/s})$           & -& -        & $3\times10^{49}$ \\ 
 Proton higher spectral cutoff & $\gamma_{p,max}$             &- & -    & $3\times 10^8$ \\ 
 Proton spectral index & $p_{p}$                              &- & -           & 2.37 \\ \hline
 Jet Power & $W_j\, (\mathrm{erg/s})$                         &$5\times 10^{45}$ & -                & -                  \\ 
 Opening angle & $\theta_{open}$                              &8 & -                & -                  \\ 
 Observing angle & $\theta_{obs}$                             &1.5 & -                & -                 \\ 
 Equilibrium factor & $A_{eq}$                                &0.1 & -                & -                 \\ 
 Number of zones  & $N$                                       &25 & -                & -                 \\ \hline
\end{tabular}
\tablefoot{Column (1) is for the SSC multi-zone model at the base of the jet; column (2) for the single-zone SSC; and column (3) for the hadronic model. We note that the models are not fitted to the data, so that these model parameters are one representation of the data and not a unique solution.
}
\end{table*}

\begin{figure}
\centering
 \resizebox{\hsize}{!}{\includegraphics[width=\textwidth]{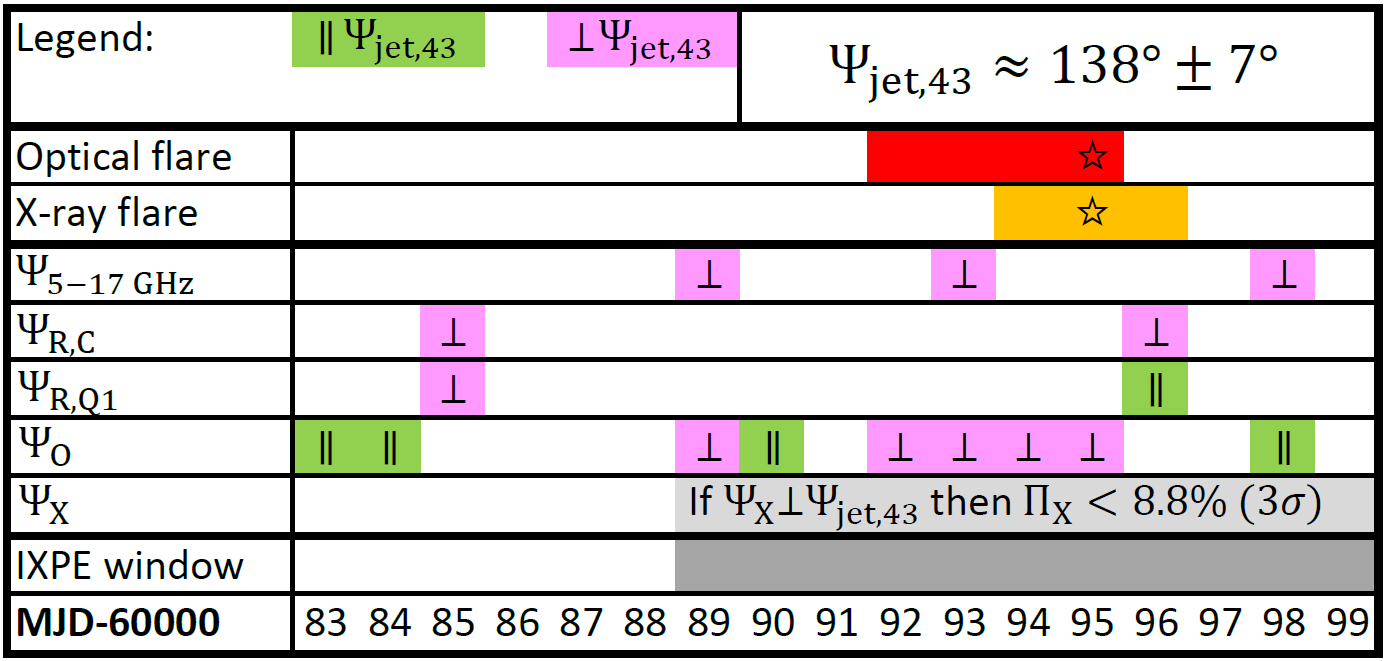}}
 \caption{Rough general timeline of the MWL observations of \blz.}
\label{fig_simplified}
\end{figure}

\section{Discussion and conclusion} \label{sec_discussion}

In this paper, we present the first X-ray polarization observation of the blazar \blz\ in a flaring optical and X-ray state, which is the first X-ray polarization observation of a purely LSP blazar in outburst. All previous IXPE observations of purely LSP blazars were performed during quiescent states (\citealt{Marshall2023_ixpe_lsp_isp}) and those of the transitional LSP-ISP blazars, BL Lacertae and S5~0716+714, were either in an LSP-quiescent state or ISP-outburst state (\citealt{Middei2023_ixpe_bllac,Peirson2023_ixpe_bllac,Marshall2023_ixpe_lsp_isp}). 

Despite the X-ray flare, we are only able to place upper limits on the X-ray polarization degree, ($\Pi_\mathrm{X} <14\%$ at 99.73\% confidence level, when considering the full, 2--8~keV, IXPE observation). We do not find any evidence for variability in either time or energy within the IXPE observing window. To obtain more constraining upper limits, we fix the X-ray polarization angle ($\Psi_\mathrm{X}$) to values motivated by the direction of the jet on the sky or the optical polarization angle ($\Psi_\mathrm{O}$). Table \ref{tab_IXPE_results} summarizes the upper limits on the X-ray polarization degree in each of the scenarios. The most constraining upper limits we find are $\Pi_\mathrm{X} <8.4\%$ and $\Pi_\mathrm{X} <8.8\%$ (at 99.73\% confidence levels), occurring when the X-ray polarization angle is assumed to be perpendicular to the constant optical polarization angle and the jet axis, respectively. However, since the optical and X-ray flares occurred in close proximity, the most likely scenario may be the latter (i.e., $\Psi_\mathrm{X} \perp \Psi_\mathrm{jet,43}$). This is because during the optical-X-ray flare $\Psi_\mathrm{O}$ was perpendicular to $\Psi_\mathrm{jet,43}$, which implies that the magnetic field lines were parallel to the jet. For hadronic synchrotron emission originating from the same region, we would then expect $\Psi_\mathrm{X} \perp \Psi_\mathrm{jet,43}$. On the other hand, inferring a likely alignment for $\Psi_\mathrm{X}$ in the SSC scenarios is more complicated, as it depends on the orientation of the seed photon region, upscattering region, and line of sight; however, if we assume that all three are roughly aligned, then SSC would generally retain the direction of $\Psi_\mathrm{O}$ and yield $\Psi_\mathrm{X} \perp \Psi_\mathrm{jet,43}$. Nevertheless, we emphasize that without a significant detection of the polarization degree, any polarization angle inference should be treated with caution.

The simultaneous optical polarization observations show variations in both the polarization degree ($\Pi_\mathrm{O}$) and angle ($\Psi_\mathrm{O}$). $\Pi_\mathrm{O}$ varied stochastically about an average value of 14.3\% with a standard deviation of 4.1\%. The most extreme change in $\Pi_\mathrm{O}$ within the IXPE observation window occurred simultaneously with a $\sim$1~magnitude optical flare (from MJD 60092 to 60095), as $\Pi_\mathrm{O}$ increased from $\sim$6\% to $\sim$20\% (see Fig.~\ref{fig_all}). Intriguingly, exactly during this flaring period, $\Psi_\mathrm{O}$ remained stable at $66.5^\circ \equiv -113.5^\circ$, which is roughly perpendicular to the jet position angle ($\Psi_\mathrm{jet,43}=-42 \pm 7$; see Sect. \ref{sec_mwl_vlbi_maps}). Outside of this 3~d flaring time window, $\Psi_\mathrm{O}$ varied rapidly from approximately parallel to the jet axis to perpendicular and vice versa; even a daily $\Delta \Psi_\mathrm{O}\sim90^\circ$ was detected (see Sect. \ref{sec_opt_obs}). In Fig.~\ref{fig_simplified} we present a rough timeline of how the MWL polarization angles of \blz\ behaved.

Meanwhile, the radio polarization degree ($\Pi_\mathrm{R}$) stayed consistently low ($<4\%$; the average was 2.2\% during the IXPE pointing with a standard deviation of 0.7\%) and the radio polarization angles ($\Psi_\mathrm{R}$), at the lower radio frequencies, were roughly perpendicular to $\Psi_\mathrm{jet,43}$ during the entire IXPE pointing. On the 43~GHz parsec-scale VLBA images of \blz\ (Fig.~\ref{fig_43GHz_VLBA}), we see a core component (C) and a second component (Q1) moving along the jet axis (i.e., northwest of the core) at an apparent superluminal motion of $18 \pm 2~c$. The flux of C at 43~GHz dominates over Q1 by a factor of $\sim$7. Hence, the point-source polarization properties in the lower-frequency radio bands mainly probe the core. Indeed, when focusing on the VLBA maps obtained close to the IXPE pointing (MJD 60085 and 60096, panels B and C of Fig.~\ref{fig_43GHz_VLBA}, respectively), we find that the polarization properties of C ($\Pi_\mathrm{R,C} \sim 3\%$ and $\Psi_\mathrm{R,C} \sim 40^\circ$, on average; see Table \ref{table_VLBA}) match those of the single-dish observations at the lower-frequency radio bands. While the lack of variability in component C near the IXPE pointing is consistent with the single-dish radio observations, it fails to explain why the optical properties of \blz\ varied so rapidly during our MWL time window. Instead, we can consider the moving component Q1 as a possible source of the optical flare. At MJD 60085, $\Psi_\mathrm{R,Q1}$ was $25^\circ$, but it changed to $132^\circ$ at MJD 60096. This behavior generally resembles that of $\Psi_\mathrm{O}$ around the same time (see Fig.~\ref{fig_simplified}). Additionally, $\Pi_\mathrm{R,Q1}$ on MJD 60085 was 1.79\%, which increased to 6.00\% on MJD 60096. This tripling of the polarization degree of Q1 coincides with the increase in $\Pi_\mathrm{O}$ during the optical flare (from 6\% to 20\%). These similarities suggest that the observed optical variability and flare originated in the moving component Q1. This is consistent with previous observations of \blz\ (\citealt{Morozova2014_fast_flare_of_2011_knot_ejection}; also see Sect. \ref{sec_intro}) and other blazars (e.g., \citealt{Marscher2010_VLBA_components_causing_flares}), where optical variability has been associated with moving parsec-scale components.

Next, we examine the timing of the optical and X-ray flares. As described in Sect. \ref{sec_x_obs_ixpe} and visible in Fig.~\ref{fig_IXPE_flux}, the X-ray flux of \blz\ measured by IXPE culminated between MJD 60094 and 60096 (peaking at MJD 60095). Meanwhile, there was a $\sim$1~magnitude optical flare which similarly peaked at MJD 60095. The lack of time delay between these flares implies that their emission regions are at least partially co-spatial. Since we have deduced that the optical flare most likely occurred in VLBA component Q1, it is likely that the X-ray flare also originated there.

Subsequently, we used the available data to constrain possible particle acceleration mechanisms within the moving component Q1, which likely caused the optical-X-ray flare. We began by considering the three commonly favored particle acceleration scenarios in blazar jets: shocks, magnetic reconnection, and turbulence. In diffusive shock acceleration, a shock front moving down the plasma jet energizes particles when they reflect back and forth across the front as they encounter magnetic irregularities (e.g., \citealt{Blandford1987_shock_acc, Marscher1985_shock_wave, sikora1994}). In magnetic reconnection scenarios, magnetic energy is efficiently converted into nonthermal particle energy at the interface between contiguous regions of parallel and antiparallel magnetic field lines (e.g., \citealt{Guo2020_reconnection_review, Zhang2020_striped_jet_mag_reconn}). These mechanisms can also occur in turbulent plasma either crossing a shock (e.g., \citealt{Marscher2014_multi_zone_shock}) or stochastically bringing together regions of opposite magnetic polarity, causing multiple reconnection events (e.g., \citealt{Comisso2018_turbulent_acc, Zhang2023_mag_reconnection_TEMZ_like}). In general, turbulence-based acceleration scenarios result in stochastic variations of the flux and polarization properties, especially at higher frequencies; this implies that their polarization angles can rotate with an arbitrary amplitude ($\Delta \Psi$), which tends to be greater when $\Pi$ is near zero. The rotation is in a random direction (clockwise or counterclockwise), usually with no relation to flaring behavior. Notably, a similar stochastic behavior can also be expected from both shock-driven and magnetic reconnection acceleration scenarios, especially if the observed radiation originates from beyond their acceleration zone, where the environment may be more turbulent. However, they both can also result in non-stochastic changes in $\Pi$ and $\Psi$, often in connection with flares, if the observed emission originates from less turbulent regions. In shocks, these changes are generally expected to be smoother, slower, and lower in amplitude compared to those that occur in magnetic reconnection (e.g., \citealt{DiGesu2022_pol_Shocks_v_MagRecon_v_Turbulence}). 

In the case of \blz, whose synchrotron emission peaks in or near the optical band, $\Pi_\mathrm{O}$ appears to behave rather stochastically. This is in agreement with the turbulent element of all three aforementioned acceleration scenarios. However, it is difficult to ascertain whether the behavior of $\Psi_\mathrm{O}$ is completely stochastic, or has some ordered nature, owing to the rather short observation window of our MWL campaign. Nevertheless, it appears that $\Psi_\mathrm{O}$ preferentially stays parallel to the jet axis, while it rotates to be roughly perpendicular to the jet axis during the optical-X-ray flare, before returning back its original orientation. If this $\Psi_\mathrm{O}$ rotation is indeed caused by an ordered component of the magnetic field, we can disfavor stochastically driven acceleration scenarios. Additionally, even in a stochastic scenario, the rather high $\Pi_\mathrm{O}$ of $\sim$15--20\% would imply a small number of emitting cells. These findings would support the shock-driven and magnetic reconnection scenarios, where the observed radiation comes from a partially ordered region.

In general, $\Psi_\mathrm{O}$ executes some rather long rotations ($\Delta \Psi_\mathrm{O}$ up to $\sim$$90^\circ$), which favors magnetic reconnection over shocks. However, since we measure the combined polarization properties of all of the components of \blz, the rotations in $\Psi_\mathrm{O}$ could also be due to an interplay between its various polarized components (e.g., \citealt{Cohen2020_interplay_causing_EVPA_rotations}). During a flare, this interplay is expected to have a weak effect, as the flaring component would be the sole dominant component. Notably, this is exactly what we have observed: during the optical-X-ray flare, $\Psi_\mathrm{O}$ remained nearly constant for three days. Such smooth and slow changes in $\Psi_\mathrm{O}$ during the flare suggests shocks. Furthermore, since $\Pi_\mathrm{O}/\Pi_\mathrm{R,Q1}$ was at least 2 during the optical-X-ray flare, the already well-established energy-stratified picture (invoked when explaining the low-energy component of blazar emission; e.g., \citealt{Liodakis2022, Marscher2024, Kouch2024_pks2155}) also applies in the case of \blz.

For most magnetic field geometries proposed for blazars, the polarization angle of the synchrotron emission of blazars ($\Psi_\mathrm{O}$ in the case of \blz) is expected to lie either parallel or perpendicular with respect to the jet position angle (e.g., \citealt{Laing1980_M87_B_parallel_to_jet, Lyutikov2005_pol_properties_of_pc_jets, Jorstad2007_pol_properties_of_pc_jets}). In the case of BLLs, it is more often parallel to the jet axis (\citealt{Hovatta2014_mojave}). In shock fronts that compress the plasma along the jet, the magnetic field lines are expected to become more perpendicular to the jet axis. In addition, a helical magnetic field is more strongly toroidal than poloidal for most viewing angles; this results in polarization angles that are aligned with the axis of a relativistic jet \citep{Lyutikov2005_pol_properties_of_pc_jets}. On the other hand, when the longitudinal components of the ambient magnetic field are amplified by shearing of the internal jet flow or by interactions with the external medium, the magnetic field lines become more parallel to the jet axis. This shearing often happens close to the edge of jets (e.g., \citealt{Laing1980_M87_B_parallel_to_jet, Pushkarev2023_B_parallel_to_jet_in_edges}). In such jets, while the passage of a shock would not amplify the ambient magnetic field, it would lead to more efficient particle acceleration in shocks that pass through the region. This would then result in a flare at higher frequencies, with polarization angles that are perpendicular to the jet axis (e.g., \citealt{Marscher2002_pol_properties_of_pc_jets}). This is precisely what we observe from \blz\ during its optical-X-ray flare, which probably took place in its moving component Q1. We therefore propose that the flare occurs in Q1 as a moving shock (which creates Q1) propagates through a segment of the jet containing a magnetic field that is parallel to the jet axis. This accelerates particles, which then cool (leading to the optical-X-ray flare) in a region that has magnetic field lines aligned parallel to the jet axis. After the flare, the field encountered by the shock becomes more turbulent, such that the shock compresses the field to partially order it perpendicular to the jet axis. This causes the polarization angle to revert to a direction parallel to the jet axis. Nevertheless, we emphasize that this interpretation may not hold if there is any significant contribution to $\Pi_\mathrm{O}$ and $\Psi_\mathrm{O}$ from any other polarized components (e.g., C).

Regarding the high-energy component of blazar emission, as described in Sect.\ref{sec_intro}, Compton scattering (either EC and SSC) or hadronic processes could be operating. As seen from the SED (top panel in Fig.~\ref{fig_SED_models}), all of the three models can adequately explain the observed data in the frequency ranges where they are relevant (see Sect.\ref{sec_models} for more details). Thus, to distinguish among them, we focus on the SPD (bottom panel in Fig.~\ref{fig_SED_models}). It is apparent that all three of the models can adequately predict $\Pi_\mathrm{O} \lesssim 14\%$. However, for the high-energy component (X-ray in the case of \blz), the SSC models predict $\Pi_\mathrm{X} < \Pi_\mathrm{O}$, while the hadronic model predicts $\Pi_\mathrm{X} \sim \Pi_\mathrm{O}$. As shown in Table \ref{tab_IXPE_results}, we obtain $\Pi_\mathrm{X} < 14\%$ and $\Pi_\mathrm{X} < 8.8\%$ (at 99.73\% confidence levels) when no assumption about $\Psi_\mathrm{X}$ is made and when the most likely alignment scenario (i.e., $\Psi_\mathrm{X} \perp \Psi_\mathrm{jet,43}$; see above for justification) is assumed, respectively. While these upper limits do not allow us to definitively distinguish between leptonic and hadronic scenarios, they strongly disfavor purely hadronic scenarios. This is consistent with all of the previous IXPE observations of LSP and ISP sources \citep{Middei2023_ixpe_bllac, Marshall2023_ixpe_lsp_isp, Peirson2023_ixpe_bllac}. Another argument against a purely hadronic scenario for \blz\ comes directly from the aforementioned conclusion that the optical and X-ray emission regions are roughly co-spatial during the flare. This conflicts with a hadronic scenario, since protons are expected to be distributed over a more extended region than electrons (\citealt{Zhang2024}) and protons need much stronger magnetic fields than electrons for efficient synchrotron cooling (\citealt{Liodakis2020}). Furthermore, a variability timescale analysis based on electron synchrotron cooling time in the observer frame $\tau_\mathrm{var}=1200 \, B^{-1.5} E^{-0.5} D^{-0.5} (1+z)$, where $\tau_\mathrm{var}$ is in seconds, $B$ is the strength of the comoving magnetic field in G, $E$ is the photon energy in keV, $D$ is the Doppler factor, and $z$ is the redshift (e.g., \citealt{Cristiani1996_varaibility_restframe_to_observed, Balokovic2016_Tvar}), reveals $\tau_\mathrm{var} \sim 13$~days for the single-zone SSC model ($B=0.028$~G and $D=30$; see Table \ref{tab_SEDmodels}) in the optical band ($E=3$~eV), which is of the same order as the observed flaring timescale of $\sim$5~days. On the other hand, in the hadronic model ($B=100$~G and $D=30$), the proton synchrotron cooling timescale\footnote{The proton cooling timescale is longer by a factor of $(m_\mathrm{p}/m_\mathrm{e})^3 \approx 6 \times 10^9$ than the electron cooling timescale (e.g., \citealt{Begelman1990_hadronic_Tvar}).} in the observer frame is $\sim$26~years in the IXPE band ($E=5$~keV). While faster variability in the hadronic model may still arise due to more rapid plasma dynamical timescales, the observed variability timescale is more difficult to accommodate under the hadronic scenario. As a result, we ultimately favor leptonic models over hadronic ones to explain the origin of the X-ray emission of \blz\ and the emission of the high-energy component of blazars as a whole; however, without a significant detection of $\Pi_\mathrm{X}$, alternative interpretations remain possible.

In the case of \blz, since the emitting region of the optical-X-ray flare is most likely in the Q1 component (i.e., on the order of parsecs from the core where the likely sources of EC seed photons are found), SSC is more likely than EC. Furthermore, when modeling the SED of \blz\ (see Sect. \ref{sec_models}), we found that SSC alone can fit the observed behavior quite well, without the need for a contribution from EC. 

To set more definitive constraints on the high-energy emission models of supermassive black hole jets, we need more observations of LSP blazars when they are in a brighter X-ray flux state or highly polarized in the optical band. Additionally, detailed modeling of the radio and optical polarization variability, as well as that of the radio morphology would be very helpful to further constrain the jet dynamics and put stronger constraints on the polarization degree under a hadronic model. BLLs, such as \blz, whose SEDs can be adequately described by SSC without the need for an EC component offer an additional advantage, as their SED modeling is rather straightforward. Furthermore, $\Pi_\mathrm{O}$ of \blz\ often exceeds $>25\%$ \citep{Raiteri2023_WEBT_opt_variability_study, Liodakis2023_circ_PD}, which makes it a prime target for future campaigns with IXPE and future higher-energy polarimeters (e.g., COSI; \citealt{Tomsick2023}).

\begin{acknowledgements}
We thank the anonymous referee for constructive suggestions
that helped improve and clarify the paper. The Imaging X-ray Polarimetry Explorer (IXPE) is a joint US and Italian mission. The US contribution is supported by the National Aeronautics and Space Administration (NASA) and led and managed by its Marshall Space Flight Center (MSFC), with industry partner Ball Aerospace (now, BAE Systems). The Italian contribution is supported by the Italian Space Agency (Agenzia Spaziale Italiana, ASI) through contract ASI-OHBI-2022-13-I.0, agreements ASI-INAF-2022-19-HH.0 and ASI-INFN-2017.13-H0, and its Space Science Data Center (SSDC) with agreements ASI-INAF-2022-14-HH.0 and ASI-INFN 2021-43-HH.0, and by the Istituto Nazionale di Astrofisica (INAF) and the Istituto Nazionale di Fisica Nucleare (INFN) in Italy. This research used data products provided by the IXPE Team (MSFC, SSDC, INAF, and INFN) and distributed with additional software tools by the High-Energy Astrophysics Science Archive Research Center (HEASARC), at NASA Goddard Space Flight Center (GSFC). This work has been partially supported by the ASI-INAF program I/004/11/4. The IAA-CSIC co-authors acknowledge financial support from the Spanish "Ministerio de Ciencia e Innovaci\'{o}n" (MCIN/AEI/ 10.13039/501100011033) through the Center of Excellence Severo Ochoa award for the Instituto de Astrof\'{i}isica de Andaluc\'{i}a-CSIC (CEX2021-001131-S), and through grants PID2019-107847RB-C44 and PID2022-139117NB-C44. The Submillimeter Array is a joint project between the Smithsonian Astrophysical Observatory and the Academia Sinica Institute of Astronomy and Astrophysics and is funded by the Smithsonian Institution and the Academia Sinica. Maunakea, the location of the SMA, is a culturally important site for the indigenous Hawai'ian people; we are privileged to study the cosmos from its summit. E.L. was supported by Academy of Finland projects 317636 and 320045. D.B. acknowledges support from the European Research Council (ERC) under the Horizon ERC Grants 2021 programme under the grant agreement No. 101040021. The research at Boston University was supported in part by National Science Foundation grant AST-2108622, NASA Fermi Guest Investigator grants 80NSSC23K1507 and 80NSSC23K1508, and NASA \textit{Swift} Guest Investigator grant 80NSSC23K1145. The Perkins Telescope Observatory, located in Flagstaff, AZ, USA, is owned and operated by Boston University. IL and SK were funded by the European Union ERC-2022-STG - BOOTES - 101076343. Views and opinions expressed are however those of the author(s) only and do not necessarily reflect those of the European Union or the European Research Council Executive Agency. Neither the European Union nor the granting authority can be held responsible for them. Some of the data are based on observations collected at the Centro Astron\'{o}mico Hispano en Andalucía (CAHA), operated jointly by Junta de Andaluc\'{i}a and Consejo Superior de Investigaciones Cient\'{i}ficas (IAA-CSIC). This work was supported by NSF grant AST-2109127. We acknowledge the use of public data from the \textit{Swift} data archive. Based on observations obtained with \textit{XMM-Newton}, an ESA science mission with instruments and contributions directly funded by ESA Member States and NASA. Partly based on observations with the 100-m telescope of the MPIfR (Max-Planck-Institut f\"ur Radioastronomie) at Effelsberg. Observations with the 100-m radio telescope at Effelsberg have received funding from the European Union’s Horizon 2020 research and innovation programme under grant agreement No 101004719 (ORP). I.L. was supported by the NASA Postdoctoral Program at the Marshall Space Flight Center, administered by Oak Ridge Associated Universities under contract with NASA. HZ is supported by NASA under award number 80GSFC21M0002. HZ's work is supported by Fermi GI program cycle 16 under the award number 22-FERMI22-0015. G.F.P. acknowledges support by the European Research Council advanced grant “M2FINDERS - Mapping Magnetic Fields with INterferometry Down to Event hoRizon Scales” (Grant No. 101018682). S. Kang, S.-S. Lee, W. Y. Cheong, S.-H. Kim, and H.-W. Jeong were supported by the National Research Foundation of Korea (NRF) grant funded by the Korea government (MIST) (2020R1A2C2009003). The KVN is a facility operated by the Korea Astronomy and Space Science Institute. The KVN operations are supported by KREONET (Korea Research Environment Open NETwork) which is managed and operated by KISTI (Korea Institute of Science and Technology Information). This work was supported by JST, the establishment of university fellowships towards the creation of science technology innovation, Grant Number JPMJFS2129. This work was supported by Japan Society for the Promotion of Science (JSPS) KAKENHI Grant Numbers JP21H01137. This work was also partially supported by Optical and Near-Infrared Astronomy Inter-University Cooperation Program from the Ministry of Education, Culture, Sports, Science and Technology (MEXT) of Japan. We are grateful to the observation and operating members of Kanata Telescope. Data from the Steward Observatory spectropolarimetric monitoring project were used. This program is supported by Fermi Guest Investigator grants NNX08AW56G, NNX09AU10G, NNX12AO93G, and NNX15AU81G. This research was partially supported by the Bulgarian National Science Fund of the Ministry of Education and Science under grants KP-06-H38/4 (2019) and KP-06-PN-68/1 (2022). The Liverpool Telescope is operated on the island of La Palma by Liverpool John Moores University in the Spanish Observatorio del Roque de los Muchachos of the Instituto de Astrofisica de Canarias with financial support from the UKRI Science and Technology Facilities Council (STFC) (ST/T00147X/1). This study makes use of VLBA data from the VLBA-BU Blazar Monitoring Program (BEAM-ME and VLBA-BU-BLAZAR; http://www.bu.edu/blazars/BEAM-ME.html), funded by NASA through the Fermi Guest Investigator Program. The VLBA is an instrument of the National Radio Astronomy Observatory. The National Radio Astronomy Observatory is a facility of the National Science Foundation operated by Associated Universities, Inc. Based on observations obtained with the Samuel Oschin Telescope 48-inch and the 60-inch Telescope at the Palomar Observatory as part of the Zwicky Transient Facility project. ZTF is supported by the National Science Foundation under Grant No. AST-2034437 and a collaboration including Caltech, IPAC, the Weizmann Institute for Science, the Oskar Klein Center at Stockholm University, the University of Maryland, Deutsches Elektronen-Synchrotron and Humboldt University, the TANGO Consortium of Taiwan, the University of Wisconsin at Milwaukee, Trinity College Dublin, Lawrence Livermore National Laboratories, and IN2P3, France. Operations are conducted by COO, IPAC, and UW. The ZTF forced-photometry service was funded under the Heising-Simons Foundation grant \#12540303 (PI: M.J.Graham). This work has made use of data from the Asteroid Terrestrial-impact Last Alert System (ATLAS) project. The Asteroid Terrestrial-impact Last Alert System (ATLAS) project is primarily funded to search for near earth asteroids through NASA grants NN12AR55G, 80NSSC18K0284, and 80NSSC18K1575; byproducts of the NEO search include images and catalogs from the survey area. This work was partially funded by Kepler/K2 grant J1944/80NSSC19K0112 and HST GO-15889, and STFC grants ST/T000198/1 and ST/S006109/1. The ATLAS science products have been made possible through the contributions of the University of Hawai'i Institute for Astronomy, the Queen’s University Belfast, the Space Telescope Science Institute, the South African Astronomical Observatory, and The Millennium Institute of Astrophysics (MAS), Chile. This publication makes use of the Mets\"ahovi Radio Observatory facilities, operated by the Aalto University, Finland. We thank Talvikki Hovatta and Tuomas Savolainen for fruitful discussions.
\end{acknowledgements}

\bibliographystyle{aa} 
\bibliography{ref} 

\begin{appendix}

\onecolumn
\section{\textit{Swift}-XRT and \textit{XMM-Newton} data} \label{appendix_Swift_XMM_spectra}
In Table \ref{tab_xrayfits_xmm_swift}, we tabulate the best-fit \textit{Swift}-XRT and \textit{XMM-Newton} results.

\begin{table*} [h!]
\caption{Best-fit \textit{Swift}-XRT and \textit{XMM-Newton} results when individually fitting the spectra.}
\label{tab_xrayfits_xmm_swift}
\begin{tabular}{llllllll}
\hline \hline
Instrument & Obs. ID & MJD & $t_\mathrm{exp}$ & Photon index & Flux & C-statistics & Theoretical C-statistics \\
\hline
\multirow{19}{*}{\textit{Swift}-XRT} & 00015977001 & 60050.2 & 1.573 & $1.33^{+0.11}_{-0.11}$ & $8.8^{+0.9}_{-0.9}$ & 106.04 & $186.43\pm16.07$ \\
 & 00015977002 & 60053.0 & 1.801 & $1.34^{+0.11}_{-0.13}$ & $7.0^{+1.0}_{-0.7}$ & 124.63 & $153.38\pm14.53$ \\
 & 00015977003 & 60060.0 & 1.281 & $1.31^{+0.10}_{-0.10}$ & $13.7^{+1.3}_{-1.2}$ & 178.81 & $205.77\pm17.51$ \\
 & 00097199001 & 60087.0 & 0.931 & $1.58^{+0.15}_{-0.14}$ & $7.6^{+1.0}_{-0.9}$ & 113.87 & $112.41\pm12.23$ \\
 & 00097199002 & 60088.0 & 0.862 & $1.55^{+0.15}_{-0.17}$ & $6.7^{+1.1}_{-0.8}$ & 84.18 & $100.31\pm11.27$ \\
 & 00097199003 & 60089.0 & 0.804 & $1.46^{+0.14}_{-0.14}$ & $8.8^{+1.3}_{-1.1}$ & 86.1 & $106.60\pm12.02$ \\
 & 00097199004 & 60090.0 & 0.969 & $1.62^{+0.15}_{-0.11}$ & $8.8^{+0.8}_{-1.2}$ & 119.44 & $142.18\pm13.86$ \\
 & 00097199005 & 60091.1 & 1.004 & $1.65^{+0.10}_{-0.19}$ & $8.9^{+1.8}_{-0.6}$ & 92.74 & $146.15\pm14.03$ \\
 & 00097199006 & 60092.1 & 0.764 & $1.35^{+0.13}_{-0.17}$ & $8.9^{+1.7}_{-0.9}$  & 92.47 & $103.63\pm11.66$ \\
 & 00097199007 & 60093.0 & 0.762 & $1.74^{+0.12}_{-0.17}$ & $12.3^{+2.1}_{-1.0}$ & 85.84 & $123.22\pm12.78$ \\
 & 00097199008 & 60094.0 & 0.782 & $1.52^{+0.11}_{-0.08}$ & $21.6^{+1.5}_{-2.0}$ & 233.57 & $227.68\pm18.69$ \\
 & 00097199009 & 60095.6 & 1.004 & $1.47^{+0.06}_{-0.09}$ & $27.7^{+2.4}_{-1.3}$ & 249.49 & $300.64\pm22.28$ \\
 & 00097199010 & 60096.2 & 0.986 & $1.29^{+0.13}_{-0.08}$ & $20.4^{+1.3}_{-2.4}$ & 174.99 & $232.06\pm18.06$ \\
 & 00097199011 & 60097.1 & 0.984 & $1.50^{+0.10}_{-0.10}$ & $15.6^{+1.4}_{-1.3}$ & 157 & $209.34\pm17.57$ \\
 & 00097199012 & 60098.1 & 0.734 & $1.44^{+0.11}_{-0.15}$ & $12.5^{+2.0}_{-1.1}$ & 107.05 & $146.10\pm14.05$ \\
 & 00097199013 & 60099.1 & 1.349 & $1.43^{+0.09}_{-0.11}$ & $11.7^{+1.3}_{-0.9}$ & 177.9 & $201.01\pm16.92$ \\
 & 00097199014 & 60100.0 & 0.892 & $1.53^{+0.14}_{-0.18}$ & $7.9^{+1.4}_{-0.9}$ & 73.04 & $94.25\pm1.18$ \\
 & 00097199015 & 60101.3 & 0.832 & $1.50^{+0.14}_{-0.14}$ & $10.7^{+1.4}_{-1.2}$ & 113.84 & $120.44\pm12.54$ \\
 & 00097199016 & 60102.2 & 1.184 & $1.56^{+0.12}_{-0.11}$ & $9.8^{+1.0}_{-1.0}$ & 115.5 & $181.46\pm16.20$ \\
 \hline
\textit{XMM-Newton} & 0920901901 & 60084.9 & 7.711 & $1.548^{+0.014}_{-0.014}$ & $8.22^{+0.12}_{-0.11}$ & 1622.47 & $1783.55\pm59.28$ \\
\hline
\end{tabular}
\tablefoot{The net exposure time ($t_\mathrm{exp}$) is given in ks, and the unabsorbed flux is given in $10^{-12}$\,erg\,cm$^{-2}$\,s$^{-1}$.}
\end{table*}

\FloatBarrier

\section{MWL data} \label{appendix_mwl_data}
In this appendix, we tabulate the MWL data of this study. The optical data are given in Tables \ref{tab_opt_data1} and \ref{tab_opt_data2}, while radio in Table \ref{tab_radio_results}. \textit{Fermi} $\gamma$-ray spectrum measurements are given in Table \ref{tab_fermi_nuFnu}. 

\begin{table} [h!]
\setlength{\tabcolsep}{1.9pt}
\centering
\caption{Measurements of brightness and linear polarization of \blz\ in optical bands via dedicated observations. \label{tab_opt_data1}}
        \begin{tabular}{c|c|c c c c c c c}
    \hline\hline 
    Ins. & Band & MJD & m & $\sigma_\mathrm{m}$ & $\Pi_\mathrm{O}$ & $\sigma_\mathrm{\Pi_\mathrm{O}}$ & $\Psi_\mathrm{O}$ & $\sigma_{\Psi_\mathrm{O}}$ \\
    \hline\hline
    \multirow{12}{*}{\rotatebox[origin=c]{90}{NOT}} & B & 60088.944 & -- & -- & 13.3\% & 0.4\% & 69.3$^\circ$ & 0.7$^\circ$ \\
     & B & 60092.926 & -- & -- & 14.2\% & 0.3\% & 72.5$^\circ$ & 0.5$^\circ$ \\
     & B & 60097.899 & -- & -- & 10.8\% & 0.3\% & 150.6$^\circ$ & 0.6$^\circ$ \\
    \cline{2-9} 
     & V & 60088.948 & -- & -- & 13.4\% & 0.2\% & 65.9$^\circ$ & 0.4$^\circ$ \\
     & V & 60092.930 & -- & -- & 13.6\% & 0.2\% & 69.0$^\circ$ & 0.3$^\circ$ \\
     & V & 60097.902 & -- & -- & 11.6\% & 0.2\% & 146.6$^\circ$ & 0.3$^\circ$ \\
     \cline{2-9} 
     & R & 60088.939 & -- & -- & 13.8\% & 0.2\% & 63.1$^\circ$ & 0.3$^\circ$ \\
     & R & 60092.903 & -- & -- & 12.1\% & 0.1\% & 64.6$^\circ$ & 0.2$^\circ$ \\
     & R & 60097.893 & -- & -- & 11.9\% & 0.1\% & 141.0$^\circ$ & 0.2$^\circ$ \\
     \cline{2-9} 
     & I & 60088.941 & -- & -- & 14.0\% & 0.2\% & 61.4$^\circ$ & 0.3$^\circ$ \\
     & I & 60092.906 & -- & -- & 11.8\% & 0.1\% & 63.0$^\circ$ & 0.2$^\circ$ \\
     & I & 60097.896 & -- & -- & 12.4\% & 0.1\% & 140.6$^\circ$ & 0.3$^\circ$ \\
    \hline\hline
    \multirow{10}{*}{\rotatebox[origin=c]{90}{T90}} & B & 60074.867 & 15.92 & 0.01 & -- & -- & -- & -- \\
     & B & 60075.988 & 15.31 & 0.01 & -- & -- & -- & -- \\
     \cline{2-9} 
     & V & 60074.869 & 15.25 & 0.01 & -- & -- & -- & -- \\
     & V & 60075.990 & 14.73 & 0.01 & -- & -- & -- & -- \\
     \cline{2-9} 
     & R & 60065.974 & 14.70 & 0.01 & -- & -- & -- & -- \\
     & R & 60067.878 & 14.13 & 0.01 & 36.3\% & 0.4\% & 89.6$^\circ$ & 0.3$^\circ$ \\
     & R & 60074.871 & 14.71 & 0.01 & -- & -- & -- & -- \\
     & R & 60075.992 & 14.21 & 0.01 & -- & -- & -- & -- \\
     \cline{2-9} 
     & I & 60074.871 & 14.01 & 0.01 & -- & -- & -- & -- \\
     & I & 60075.993 & 13.55 & 0.01 & -- & -- & -- & -- \\
    \hline\hline
    \multirow{16}{*}{\rotatebox[origin=c]{90}{LX-200}} & R & 60087.898 & 14.33 & 0.02 & 6.5\% & 0.5\% & 113.5$^\circ$ & 2.1$^\circ$ \\
     & R & 60088.887 & 14.35 & 0.02 & 17.87\% & 0.48\% & 67.8$^\circ$ & 0.8$^\circ$ \\
     & R & 60089.913 & 14.30 & 0.02 & 12.70\% & 0.56\% & 148.8$^\circ$ & 1.3$^\circ$ \\
     & R & 60090.900 & 14.08 & 0.02 & 16.09\% & 0.77\% & 108.1$^\circ$ & 1.4$^\circ$ \\
     & R & 60091.901 & 14.45 & 0.02 & 6.15\% & 0.60\% & 61.7$^\circ$ & 2.8$^\circ$ \\
     & R & 60092.906 & 14.12 & 0.02 & 13.83\% & 0.57\% & 67.3$^\circ$ & 1.2$^\circ$ \\
     & R & 60094.898 & 13.23 & 0.01 & 19.21\% & 0.36\% & 65.8$^\circ$ & 0.6$^\circ$ \\
     & R & 60094.917 & 13.24 & 0.02 & 19.28\% & 0.51\% & 69.0$^\circ$ & 0.8$^\circ$ \\
     & R & 60094.929 & 13.21 & 0.02 & 19.89\% & 0.44\% & 72.5$^\circ$ & 0.7$^\circ$ \\
     \cline{2-9} 
     & I & 60087.910 & 13.69 & 0.02 & -- & -- & -- & -- \\
     & I & 60088.903 & 13.65 & 0.01 & -- & -- & -- & -- \\
     & I & 60089.944 & 13.63 & 0.03 & -- & -- & -- & -- \\
     & I & 60090.886 & 13.42 & 0.02 & -- & -- & -- & -- \\
     & I & 60091.913 & 13.80 & 0.01 & -- & -- & -- & -- \\
     & I & 60092.917 & 13.48 & 0.01 & -- & -- & -- & -- \\
     & I & 60094.907 & 12.64 & 0.01 & -- & -- & -- & -- \\
    \hline\hline
    \multirow{6}{*}{\rotatebox[origin=c]{90}{RoboPol}} & R & 60082.835 & 14.28 & 0.02 & 13.98\% & 0.12\% & 149.0$^\circ$ & 0.3$^\circ$ \\
    & R & 60082.842 & 14.29 & 0.02 & 13.22\% & 0.16\% & 150.1$^\circ$ & 0.4$^\circ$ \\
    & R & 60083.802 & 14.13 & 0.02 & 10.40\% & 0.10\% & 143.0$^\circ$ & 0.3$^\circ$ \\
    & R & 60104.818 & 14.12 & 0.02 & 13.48\% & 0.10\% & 154.2$^\circ$ & 0.2$^\circ$ \\
    & R & 60105.811 & 13.95 & 0.02 & 27.09\% & 0.10\% & 176.5$^\circ$ & 0.1$^\circ$ \\
    & R & 60106.824 & 13.87 & 0.02 & 17.02\% & 0.09\% & 164.5$^\circ$ & 0.2$^\circ$ \\
    \hline\hline
    \end{tabular}
    \tablefoot{``Ins.'' represents the instrument or telescope facility employed (see Sect. \ref{sec_opt_obs}). ``Band'' gives the optical band. m and $\sigma_\mathrm{m}$ refer to the brightness and its error in magnitudes, respectively. $\Pi_\mathrm{O}$ and $\sigma_\mathrm{\Pi_\mathrm{O}}$ give the optical polarization degree and its error, respectively. $\Psi_\mathrm{O}$ and $\sigma_{\Psi_\mathrm{O}}$ represent the optical polarization angle and its error, respectively. All errors are $1\sigma$ Gaussian. For brevity, in case of several equivalent (within $1\sigma$) brightness entries per one night, only one one of them is shown.}
\end{table}

\begin{table}
\setlength{\tabcolsep}{1.9pt}
\centering
\caption{Measurements of brightness of \blz\ in the optical band via all-sky surveys. \label{tab_opt_data2}}
        \begin{tabular}{c|c|c c c c c c c}
    \hline\hline 
    Survey & Band & MJD & m & $\sigma_\mathrm{m}$ \\
    \hline\hline
    \multirow{4}{*}{\rotatebox[origin=c]{90}{ZTF}} & r & 60082.199 & 14.51 & 0.02 \\
    & r & 60084.197 & 14.26 & 0.02 \\
    & r & 60097.221 & 13.71 & 0.02 \\
    & r & 60099.240 & 14.57 & 0.02 \\
    \hline\hline
    \multirow{11}{*}{\rotatebox[origin=c]{90}{ATLAS}} & o & 60082.325 & 14.59 & 0.01 \\
    & o & 60084.291 & 14.36 & 0.02 \\
    & o & 60086.262 & 14.77 & 0.03 \\
    & o & 60088.274 & 14.11 & 0.03 \\
    & o & 60090.297 & 14.67 & 0.02 \\
    & o & 60092.295 & 14.69 & 0.03 \\
    & o & 60094.282 & 13.72 & 0.01 \\
    & o & 60096.265 & 13.88 & 0.02 \\
    & o & 60098.295 & 14.20 & 0.03 \\
    & o & 60100.284 & 14.99 & 0.06 \\
    & o & 60102.263 & 15.04 & 0.02 \\
    \hline\hline
    \end{tabular}
    \tablefoot{``Survey'' represents the all-sky survey used (see Sect. \ref{sec_opt_obs}). ``Band'' gives the optical band of the measurements. m and $\sigma_\mathrm{m}$ refer to the brightness and its $1\sigma$ Gaussian error in magnitudes, respectively.}
\end{table}

\begin{table}
\setlength{\tabcolsep}{1.9pt}
\centering
\caption{Linear polarization measurements of \blz\ in radio bands.\label{tab_radio_results}}
        \begin{tabular}{l l l l l}
        \hline\hline
        Program & $\nu$ (GHz) & MJD & $\Pi_\mathrm{R}$ & $\Psi_\mathrm{R}$  \\
        \hline
    QUIVER & 4.8 & 60093.5 & $(2.43 \pm 0.47)\%$ & $60.0^\circ \pm 2.4^\circ$ \\
    QUIVER & 4.8 & 60097.7 & $(1.40 \pm 0.54)\%$ & $61.9^\circ \pm 3.1^\circ$ \\
    QUIVER & 10 & 60088.8 & $(0.68 \pm 0.15)\%$ & $61.0^\circ \pm 6.5^\circ$ \\
    QUIVER & 10 & 60093.5 & $(1.65 \pm 0.08)\%$ & $55.8^\circ \pm 2.2^\circ$ \\
    QUIVER & 10 & 60097.7 & $(1.08 \pm 0.09)\%$ & $57.9^\circ \pm 2.3^\circ$ \\
    QUIVER & 17 & 60093.5 & $(2.53 \pm 0.22)\%$ & $37.0^\circ \pm 2.2^\circ$ \\
    QUIVER & 17 & 60097.7 & $(1.92 \pm 0.28)\%$ & $50.1^\circ \pm 2.2^\circ$ \\
    SMAPOL & 225.5 & 60094.2 & $(3.16 \pm 0.39)\%$ & $115.73^\circ \pm 1.99^\circ$ \\
    SMAPOL & 225.5 & 60099.3 & $(4.05 \pm 0.50)\%$ & $115.90^\circ \pm 1.97^\circ$ \\
    \hline
        \end{tabular}
\end{table}

\begin{table}
\setlength{\tabcolsep}{1.9pt}
\centering
\caption{\textit{Fermi} $\gamma$-ray spectrum in the 0.1-100 GeV band.\label{tab_fermi_nuFnu}}
        \begin{tabular}{c|cc}
        \hline\hline
        Frequency ($\nu$) & $\nu F_\nu$ & Error of $\nu F_\nu$ \\
    (1) & (2) & (3) \\
    \hline
    22.7 & 7.1 & 1.1 \\
    23.1 & 11.1 & 1.3 \\
    23.5 & 8.2 & 1.5 \\
    24.0 & 5.2 & 2.1 \\
    24.4 & $<$~2.7 & -- \\
    24.8 & $<$~5.4 & -- \\
    25.3 & $<$~9.8 & -- \\
    \hline
        \end{tabular}
    \tablefoot{Column (1) gives frequency in log(Hz) units, column (2) gives the frequency multiplied by the flux density per frequency ($\rm \nu{F}\nu$) in units of 10$^{-11}$ erg s$^{-1}$ cm$^{-2}$, and column (3) gives the error of the values in column (2).}
\end{table}

\end{appendix}

\end{document}